\documentclass[10pt,twocolumn,journal]{IEEEtran}
\usepackage{latexsym}
\usepackage{amsfonts}
\usepackage{amsbsy}
\usepackage{amsmath,amssymb}
\usepackage{times}
\usepackage{graphicx}
\usepackage{enumerate}
\usepackage[usenames]{color}
\usepackage[dvips]{pstcol}
\title {\Huge Low-Complexity Variable Forgetting Factor Techniques for RLS Algorithms in Interference Rejection Applications}
\author{Yunlong Cai and Rodrigo C. de Lamare
\thanks{Y. Cai is with Department of
Information Science and Electronic Engineering, Zhejiang University,
Hangzhou 310027, China (e-mail: ylcai@zju.edu.cn).

R. C. de Lamare
is with the Communications Research Group, Department of
Electronics, University of York, YO10 5DD York, U.K. (e-mail:
rcdl500@ohm.york.ac.uk).

This work is supported by the Fundamental Research Funds for the Central Universities and
the NSF of China
under grant $61101103$.
} }

\begin{document}
\maketitle \thispagestyle{empty}
\begin{abstract}

\\
%In this work, we propose a low-complexity variable forgetting factor
%(VFF) mechanism for recursive least square (RLS) algorithms in
%interference suppression applications. The proposed VFF mechanism
%employs an updated component relating to the time average of the
%error correlation to automatically adjust the forgetting factor in
%order to ensure fast convergence and good tracking of the
%interference and the channel. Convergence and tracking analyses are
%carried out and analytical expressions for predicting the mean
%squared error of the proposed adaptation technique are obtained.
%Simulation results for a  direct-sequence code-division multiple
%access (DS-CDMA) system are presented for nonstationary environments and show
%that the proposed VFF mechanism achieves superior performance to
%previously reported methods at a reduced complexity.
In this work, we propose a low-complexity variable forgetting factor (VFF) mechanism for recursive least square (RLS) algorithms in interference suppression applications. The proposed VFF mechanism employs an updated component related to the time average of the error correlation to automatically adjust the forgetting factor in order to
ensure fast convergence and good tracking of the interference and the channel. Convergence and tracking analyses are carried out and analytical expressions for predicting the mean squared error of the proposed adaptation technique are obtained. Simulation results for a direct-sequence code-division multiple access (DS-CDMA) system are presented in nonstationary environments and show that the proposed VFF mechanism achieves superior performance to previously reported methods at a reduced complexity.

\emph{Index Terms}--- Interference suppression, adaptive receivers,
recursive least squares algorithm, variable forgetting factor
mechanism, time-varying channels
\end{abstract}
%\begin{keywords}
%Blind multiuser detection, adaptive filtering, variable forgetting factor mechanisms
%\end{keywords}
%\IEEEpeerreviewmaketitle
%

\section{Introduction}

Wireless communication channels are dynamic by nature and present
significant challenges to the design of receivers
\cite{tse,rappaport}. In particular, {  the estimation of the
parameters of the channels and the receive filters of wireless
receivers has received significant attention in the past years
%
%is a problem that has been
%around for many years
and is key to the performance of wireless systems} \cite{poor}. In
such nonstationary environments, {  specifically for the slow
time-varying channels,} adaptive techniques are of fundamental
importance and encounter applications in many wireless transmission
schemes such as multi-input multi-output (MIMO)
\cite{besselj1}-\cite{rcdl_mimo}, orthogonal-frequency division
multiplexing (OFDM) \cite{stuber} and spread spectrum systems
\cite{Woodward}-\cite{rcdl4}.

The most common adaptive estimation techniques are the stochastic
gradient (SG) and recursive least squares (RLS) algorithms
\cite{Haykin}.
%Despite the linear computational complexity of SG
%techniques, e.g., least mean square (LMS) algorithms, their
%performance is typically poor in correlated channels and hostile
%environments. For this reason, it is preferable to implement
%adaptive receivers with RLS algorithms due to the fast convergence
%and better steady-state performance.
{  The works in \cite{poor}, \cite{Haykin} have considered standard
SG algorithms with fixed step sizes to implement the  receiver. The
authors in \cite{chambers}, \cite{RKwong} have proposed some
variable step-size schemes to accelerate the convergence speed of
the SG based adaptive filters. The RLS algorithm is considered as
one of the fastest and most effective methods for adaptive
implementation \cite{Haykin}. }
In nonstationary wireless
environments in which users often enter and exit the system, it is
impractical to compute a predetermined value for the forgetting
factor. Therefore, the RLS algorithm needs to be modified in order
to yield satisfactory performance in time-varying environments.
%
%In
%this regard, one promising technique that has been proposed is to
%employ variable forgetting factor (VFF) mechanisms to adjust the
%forgetting factor automatically \cite{Haykin}-\cite{Leung2}. The
%classic VFF mechanism is the gradient-based variable forgetting
%factor (GVFF) algorithm which was proposed in \cite{Haykin}. In this
%work, besides the recursive expressions to adapt the receive vector,
%another SG recursion is used to control the forgetting factor, where
%the gradient with respect to the forgetting factor  is obtained
%based on the instantaneous squared error cost function.
%
In this regard, one promising technique that has been considered is to employ variable forgetting factor (VFF) mechanisms to adjust the forgetting factor automatically \cite{Haykin}-\cite{Leung2}. The classic VFF mechanism is the gradient-based variable forgetting factor (GVFF) algorithm which has been proposed in \cite{Haykin}. The work in \cite{Haykin}, besides the recursive expressions to adapt the receive vector, uses another SG recursion to control the forgetting factor where the gradient with respect to the forgetting factor is obtained based on the instantaneous squared error cost function.
In \cite{ssong3}, the authors proposed a modified variable
forgetting factor mechanism using the error criterion with noise
variance weighting for frequency selective fading channel
estimation. A VFF mechanism based on the Gauss Newton approach was
proposed in \cite{ssong1} to improve the fast time-varying parameter
estimation. In \cite{Leung1,Leung2}, a modified GVFF mechanism
based on the gradient of the mean squared error (MSE) rather than on
the gradient of the instantaneous squared error was investigated.
Another approach is to perturb the covariance matrix whenever the
change is detected \cite{djpark}, however, the algorithm becomes
sensitive to disturbance and noise. The existing VFF mechanisms are
mostly derived from the architecture of the classic GVFF algorithm
\cite{Haykin}, and {  the computational} complexity of which is
proportional to the receive filter length.

%In this work, we propose a low-complexity VFF mechanism for adaptive
%RLS algorithms applied to linear interference suppression in
%direct-sequence code-division multiple access (DS-CDMA) systems. The
%proposed VFF mechanism employs an updated component relating to the
%time average of the error correlation to automatically adjust the
%forgetting factor in order to ensure good tracking of the
%interference and the channel. We refer to the proposed VFF scheme as
%correlated time-averaged variable forgetting factor (CTVFF)
%mechanism. Convergence and tracking analyses are carried out and
%analytical expressions for predicting the mean squared error of the
%proposed adaptation technique are obtained. It should be noted that
%the proposed scheme is general and can be applied to any wireless
%systems with the RLS algorithm. Simulation results are presented for
%nonstationary environments and show that the proposed VFF mechanism
%achieves superior performance to previously reported methods at a
%reduced complexity.
In this work, we propose a low-complexity VFF mechanism for adaptive RLS algorithms applied to linear interference suppression in direct-sequence code-division multiple access (DS-CDMA) systems. The proposed VFF mechanism employs an updated component related to the time average of the error correlation to automatically adjust the forgetting factor in order to ensure good tracking of the interference and the channel. We refer to the proposed VFF scheme as correlated time-averaged variable forgetting factor (CTVFF) mechanism. Convergence and tracking analyses are carried out and analytical expressions for predicting the mean squared error of the proposed
adaptation technique are obtained. It should be noted that the proposed scheme is general and can be applied to any wireless systems with the RLS algorithm. Therefore, the scheme can be very useful in parameter estimation applications to multi-antenna and OFDM systems.

The paper is structured as follows. Section \ref{Section2:system}
briefly describes the DS-CDMA system model and the design of linear
receivers. The adaptive RLS algorithm and the existing gradient-based VFF
%
%classic adaptive GVFF
mechanisms are reviewed in {  Section} \ref{Section3:RLSCCM}. The
proposed CTVFF mechanism is described in {  Section}
\ref{Section4:proposedVFF}. Convergence and tracking analyses of the
resulting algorithm and the analytical formulas to predict the
steady-state MSE and the steady-state mean value of the proposed
CTVFF mechanism are developed in {  Section}
\ref{Section5:analysis}. The simulation results are presented in {
Section} \ref{Section6:simulations}. Finally, {  Section}
\ref{Section7:conclusion} draws the conclusions.

%%%%%%%%%%%%%%%%%%%%%%%%%%%%%%%%%%%%%%%%%%%%%%%%%%%%%%%%%

%%%%%%%%%%%%%%%%%%%%%%%%%%%%%%%%%%%%%%%%%%%%%%%%%%%%%%%

\section{DS-CDMA System Model and Design of Linear MMSE Receivers}
\label{Section2:system}

Detecting a desired signal in DS-CDMA systems requires processing
the received signal in order to mitigate the interference and the
noise at the receiver. The major source of interference in DS-CDMA
systems is multiuser interference (MUI), which arises due to the
fact that users communicate through the same physical channel with
nonorthogonal signals. Multiuser detection has been proposed as a
means to suppress MUI, increasing the capacity and the performance
of CDMA systems \cite{verdu1}-\cite{rcdl4}. The linear minimum mean
squared error (MMSE) receiver implemented with an adaptive algorithm
is one of the most prominent techniques since it only requires the
timing of the desired user and a training sequence.  In this
work, we focus on adaptive linear receivers as extensions to other
receivers are straightforward.

Let us consider the downlink of an uncoded synchronous binary phase-shift keying (BPSK) DS-CDMA system
with $K$ users, $N$ chips per symbol and $L_{p}$ propagation
paths. A synchronous model is assumed for simplicity since it
captures most of the features of more realistic asynchronous
models with small to moderate delay spreads.
%The modulation is
%assumed to have constant modulus.
Let us assume that the signal
has been demodulated at the mobile user, the channel is constant
during each symbol and the receiver is perfectly synchronized with
the main channel path. The received signal after filtering by a
chip-pulse matched filter and sampled at chip rate yields an
$M$-dimensional received vector at time $i$
%\begin{equation}
%\begin{split}
%{\boldsymbol r}(i) & = \sum_{k=1}^{K} A_{k}(i)  {b}_{k}(i)
%{\boldsymbol C}_{k} {\boldsymbol h}(i) +
% {\boldsymbol{\eta}}_k(i) + {\boldsymbol n}(i), \label{recsignal}
%\end{split}
%\end{equation}
\begin{equation}
\begin{split}
\mathbf {r}(i) & =  \sum_{k=1}^{K}\big(A_{k}b_{k}(i) \mathbf {C}_{k}\mathbf {h}(i)+\mbox{\boldmath$\eta$}_{k}(i)\big)+ \mathbf {n}(i), \label{recsignal}
\end{split}
\end{equation}
where $M=N+L_{p}-1$, $\mathbf{n}(i) = [n_{1}(i)
~\ldots~n_{M}(i)]^{T}$ is the complex Gaussian noise vector with
zero mean and $E[\mathbf{n}(i)\mathbf{n}^{H}(i)] =
\sigma^{2}\mathbf{I}$ whose components are independent and identically distributed, where $(.)^{T}$ and $(.)^{H}$ denote transpose and Hermitian
transpose, respectively, and $E[.]$ stands for expected value.
 The
user symbols are denoted by ${b}_{k}(i)$, where we assume that the symbols are independent and identically distributed random variables with equal
probability from the set $ \{\pm1\}$. The amplitude of user $k$
is $A_{k}$, and
%and $\mbox{\boldmath$\eta$}_{k}(i)$ is the intersymbol
%interference  for user $k$.
the signature of user $k$ is
represented by $\mathbf{p}_{k} = [a_{k}(1) \ldots
a_{k}(N)]^{T}$. The $M\times L_{p}$ constraint matrix $\mathbf{
C}_{k}$ that contains one-chip shifted versions of the signature
sequence for user $k$ and the $L_{p}\times 1$ vector $\mathbf{
h}(i)$ with the multipath components are described by
\begin{equation}
\mathbf{C}_{k} = \left[\begin{array}{c c c }
a_{k}(1) &  & \mathbf{ 0} \\
\vdots & \ddots & a_{k}(1)  \\
a_{k}(N) &  & \vdots \\
\mathbf{ 0} & \ddots & a_{k}(N)  \\
 \end{array}\right],
 \mathbf{ h}(i)=\left[\begin{array}{c} {h}_{0}(i)
\\ \vdots \\ {h}_{L_{p}-1}(i)\\  \end{array}\right],
\end{equation}
where the $M\times 1$ vector $\mathbf{C}_{k}\mathbf{h}(i)$ denotes the effective spreading code.
%The MAI comes from the non-orthogonality between the received
%signature sequences. {  The ISI originates from the multipath
%propagation effects of the channel, depends on the length of the
%channel response and how it is related to the length of the chip
%sequence. We define $L_s$ as the ISI span, i.e., the number of
%symbols affected by the channel. For $L_{p}=1,~ L_{s}=1$ (no ISI),
%for $1<L_{p}\leq N, L_{s}=3$, for $N <L_{p}\leq 2N, L_{s}=5$, and so
%on.} At time instant $i$ we will have ISI coming not only from the
%previous time instants but also from the next symbols.
 The vector $\mbox{\boldmath$\eta$}_{k}(i)$ denotes the  intersymbol
interference (ISI)  for user $k$, here we
express the ISI vector in a general form that is given by
 $\mbox{\boldmath$\eta$}_{k}(i)=A_{k}b_{k}(i-1)\mathbf{H}^{p}\mathbf{p}_{k}+A_{k}b_{k}(i+1)\mathbf{H}^{s}\mathbf{p}_{k}$, where the $M\times N$ matrices $\mathbf{H}^{p}$ and $\mathbf{H}^{s}$ account for the
ISI from previous and subsequent symbols, respectively, and  can be given as follows
\begin{equation*}
\mathbf{H}^{p} = \left[\begin{array}{c c c  c c}
0 & \ldots & h_{L_{p}-1}(i-1) & \ldots   & h_{1}(i-1) \\
0 &  &  &  \ddots & \vdots   \\
\vdots &  & & & h_{L_{p}-1}(i-1) \\
 0 & & & & \vdots\\
0 & 0 & \ldots & 0& 0 \\
 \end{array}\right],
% \mathbf{H}_{2}(i)=\left[\begin{array}{c} {h}_{0}(i)
%\\ \vdots \\ {h}_{L_{p}-1}(i)\\  \end{array}\right].
\end{equation*}
\begin{equation}
\mathbf{H}^{s} = \left[\begin{array}{c c c  c c}
0 & 0 &\ldots & 0  & 0 \\
\vdots &  &  &   & 0   \\
h_{0}(i+1) &  & & &\vdots \\
 \vdots & \ddots & & & 0\\
h_{L_{p}-2}(i+1) & \ldots & h_{0}(i+1) & \ldots & 0 \\
 \end{array}\right].
% \mathbf{H}_{2}(i)=\left[\begin{array}{c} {h}_{0}(i)
%\\ \vdots \\ {h}_{L_{p}-1}(i)\\  \end{array}\right].
\end{equation}

The linear MMSE receiver design is equivalent to determining a
finite impulse response (FIR) filter $\mathbf{w}_{k}(i)$ with $M$
coefficients that provide an estimate of the desired symbol as
follows
\begin{equation}
z_k(i) = \mathbf{w}_{k}^{H}(i)\mathbf{r}(i),
\label{symbol}
\end{equation}
where the detected symbol is given by $\hat{b}_{k}(i) =
\textrm {sign} \{\Re[\mathbf{w}_{k}^{H}(i)\mathbf{r}(i)]\}$,
%where ${\rm
%Q}(\cdot)$ is a function that performs the detection according to
%the constellation employed.
where the operator $\Re[.]$ retains the real part of the
argument and $\textrm {sign} \{.\}$ is the signum function.

The design of the receive filter $\mathbf{w}_{k}(i)$ is based
on the optimization of the MSE cost function
\begin{equation}
\bar{J}_{MSE}(\mathbf{w}_{k}(i)) = E\Big[|b_{k}(i)-\mathbf{
w}_{k}^{H}(i)\mathbf{r}(i)|^{2}\Big]. \label{eq:MSEcost}
\end{equation}
By minimizing (\ref{eq:MSEcost}), the MMSE receive filter is given
by \cite{verdu1}
\begin{equation}
\mathbf{w}_{0}=\mathbf{\bar{R}}^{-1} \mathbf{s},
\label{eq:MMSEEXPRESSION}
\end{equation}
%%%%%%%%%%%%%%%%%%%%%%%%%%%%%%%%%%%%%%%%%%%%%%%
%By substituting (\ref{eq:MMSEEXPRESSION}) into (\ref{eq:MSEcost}), we have
where
$\mathbf{\bar{R}}=E[\mathbf{r}(i)\mathbf{r}^{H}(i)]=\sum^{K}_{k^{'}=1}A^2_{k^{'}}\mathbf{C}_{k^{'}}\mathbf{h}(i)\mathbf{h}^{H}(i)\mathbf{C}^{H}_{k^{'}}+\sum^{K}_{k^{'}=1}A^2_{k^{'}}\mathbf{H}^{p}\mathbf{p}_{k^{'}}\mathbf{p}^{H}_{k^{'}}\mathbf{H}^{pH}+\sum^{K}_{k^{'}=1}A^2_{k^{'}}\mathbf{H}^{s}\mathbf{p}_{k^{'}}\mathbf{p}^{H}_{k^{'}}\mathbf{H}^{sH}+\sigma^2\mathbf{I}$
and ${\mathbf s}
=E[b^{*}_{k}(i)\mathbf{r}(i)]=A_{k}\mathbf{C}_{k}\mathbf{h}(i)$.
%In this work, we set the power of the desired user to be unit, $A_{k}=1$.
The minimum value of the mean squared error is given by
\begin{equation}
\xi_{\rm
min}=1-A^{2}_{k}\mathbf{h}^{H}(i)\mathbf{C}^{H}_{k}\mathbf{\bar{R}}^{-1}\mathbf{C}_{k}\mathbf{h}(i).
\end{equation}

\section{ ADAPTIVE RLS ALGORITHMS AND PROBLEM STATEMENT}
\label{Section3:RLSCCM}

In this section, we first describe the adaptive RLS algorithm to
estimate the parameters of the linear MMSE receiver for multipath
channels using a least-squares (LS) approach. Then, the  existing
gradient-based  VFF mechanisms  are introduced.

\subsection{Adaptive  Receiver}

Let us consider the time-averaged cost function
\begin{equation}
J_{LS}(i)=\sum^{i}_{n=1}\lambda^{i-n}|b_{k}(n)-\mathbf{
w}_{k}^{H}(i)\mathbf{r}(n)|^{2}\label{eq:lscriterion}
\end{equation}
where $\lambda$ denotes the forgetting factor.
% and we assume that the power of the desired user is unit.
By taking the
gradient of (\ref{eq:lscriterion}) with respect to $\mathbf{w}^{*}_{k}(i)$ and setting it to zero,
after further mathematical manipulations we have the adaptive RLS algorithm as follows \cite{Haykin}
\begin{equation}
\mathbf{w}_{k}(i)=\mathbf{w}_{k}(i-1)+\mathbf{k}(i)e^{*}(i),\label{eq:filterw}
\end{equation}
where
\begin{equation}
\mathbf{k}(i)=\frac{\mathbf{R}^{-1}(i-1)\mathbf{r}(i)}{\lambda+\mathbf{r}^{H}(i)\mathbf{R}^{-1}(i-1)\mathbf{r}(i)}\label{eq:k}
\end{equation}
and
\begin{equation}
e(i)=b_{k}(i)-\mathbf{w}^{H}_{k}(i-1)\mathbf{r}(i),
\end{equation}
and the estimate $\mathbf{R}^{-1}(i)$ is updated by
\begin{equation}
\mathbf{R}^{-1}(i)=\lambda^{-1}\mathbf{R}^{-1}(i-1)-\lambda^{-1}\mathbf{k}(i)\mathbf{r}^{H}(i)\mathbf{R}^{-1}(i-1).\label{eq:filter0}
\end{equation}
The adaptive algorithm is implemented by using
(\ref{eq:filterw})-(\ref{eq:filter0}) with appropriate initial
values $\mathbf{R}^{-1}(0)$ and $\mathbf{w}_{k}(0)$. The algorithm
has been devised to start its operation in the training (TR) mode,
and then to switch to the decision-directed (DD) mode. The problem
we are interested in solving is how to devise a cost-effective
mechanism to adjust $\lambda$, which is a key factor affecting the
performance of RLS-based algorithms and the receivers.

\subsection{ Gradient-based Mechanisms }

\begin{figure}[!hhh]
\centering \scalebox{0.75}{\includegraphics{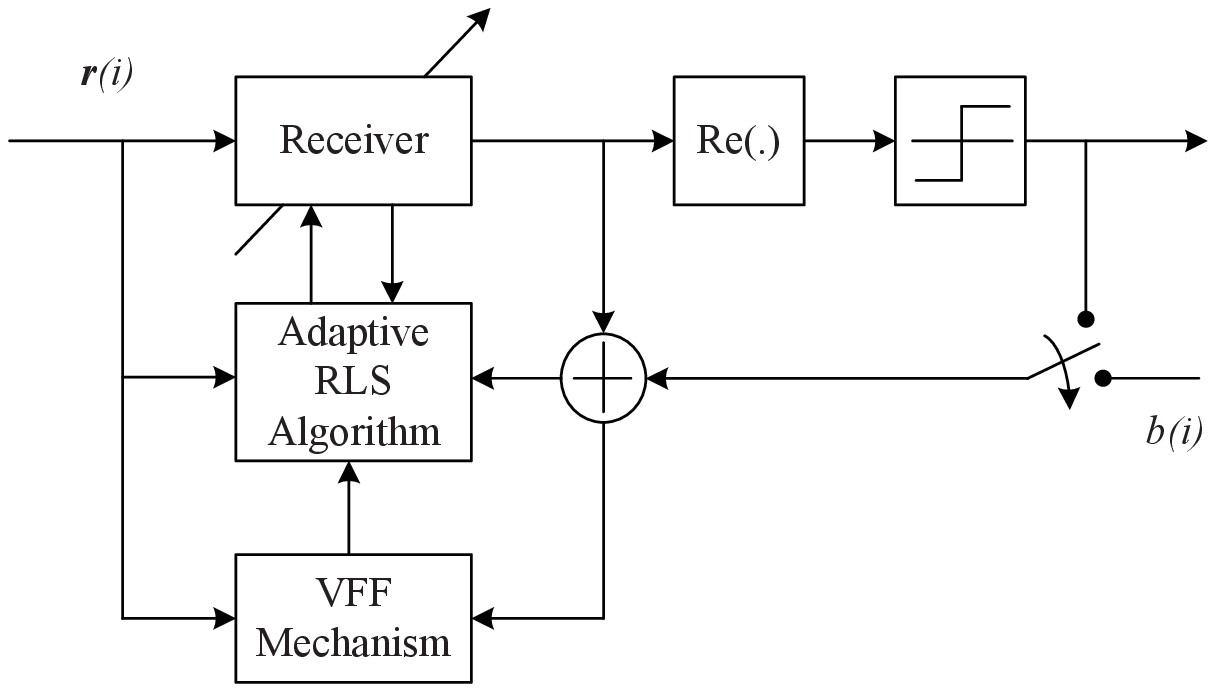}}
\vspace{-0.95em}\caption{Structure of the adaptive  receiver with
the variable forgetting factor mechanism } \label{fig:block}
\end{figure}

{  In order to adjust the forgetting factor automatically, the
adaptive rule for the classic   GVFF mechanism is derived by taking
the gradient of the instantaneous cost function $|b_{k}(i)-\mathbf{
w}_{k}^{H}(i)\mathbf{r}(i)|^2$ with respect to the variable
forgetting factor $\lambda(i)$ \cite{Haykin}. The adaptive
expressions are given by
%we have

\begin{equation}
\lambda(i)=\Big[\lambda(i-1)+\mu\Re\big[\frac{\partial\mathbf{w}^{H}_{k}(i-1)}{\partial \lambda}\mathbf{r}(i)e^{*}(i) \big] \Big]^{\lambda^{+}}_{\lambda^{-}},
\end{equation}

\begin{equation}
\frac{\partial\mathbf{w}_{k}(i)}{\partial \lambda}=\big(\mathbf{I}-\mathbf{k}(i)\mathbf{r}^{H}(i) \big)\frac{\partial\mathbf{w}_{k}(i-1)}{\partial \lambda}+\frac{\partial\mathbf{R}^{-1}(i)}{\partial \lambda}\mathbf{r}(i)e^{*}(i),
\end{equation}

\begin{equation}
\begin{split}
\frac{\partial\mathbf{R}^{-1}(i)}{\partial \lambda}&=\lambda^{-1}(i)\big(\mathbf{I}-\mathbf{k}(i)\mathbf{r}^{H}(i) \big)\frac{\partial\mathbf{R}^{-1}(i-1)}{\partial \lambda}\big(\mathbf{I}-\mathbf{r}(i)\mathbf{k}^{H}(i) \big)+
\lambda^{-1}(i)\mathbf{k}(i)\mathbf{k}^{H}(i)-\lambda^{-1}(i)\mathbf{R}^{-1}(i),\label{eq:gvff3}
\end{split}
\end{equation}
where $[.]^{\lambda^{+}}_{\lambda^{-}}$ denotes the truncation to the limits
of the range $[\lambda^{-}, \lambda^{+}]$, $\mu$ denotes a step-size.
%
%According to \cite{Haykin}, the upper level of truncation, $\lambda^{+}$, plays a relatively
%insignificant role, we can set it equal to a positive value which is less than but close to $1$.
%The lower level of truncation, $\lambda^{-}$ that ensures the stability plays a more important role and should be determined by simulations.
%Here, a new quantity $\frac{\partial\mathbf{w}_{k}(i)}{\partial \lambda}$ is introduced, the updated equation of which is given by
%\begin{equation}
%\frac{\partial\mathbf{w}_{k}(i)}{\partial \lambda}=\big(\mathbf{I}-\mathbf{k}(i)\mathbf{r}^{H}(i) \big)\frac{\partial\mathbf{w}_{k}(i-1)}{\partial \lambda}+\frac{\partial\mathbf{R}^{-1}(i)}{\partial \lambda}\mathbf{r}(i)e^{*}(i),
%\end{equation}
%where $\frac{\partial\mathbf{R}^{-1}(i)}{\partial \lambda}$ is updated by
%\begin{equation}
%\begin{split}
%\frac{\partial\mathbf{R}^{-1}(i)}{\partial \lambda}&=\lambda^{-1}(i)\big(\mathbf{I}-\mathbf{k}(i)\mathbf{r}^{H}(i) \big)\frac{\partial\mathbf{R}^{-1}(i-1)}{\partial \lambda}\big(\mathbf{I}-\mathbf{r}(i)\mathbf{k}^{H}(i) \big)+
%\lambda^{-1}(i)\mathbf{k}(i)\mathbf{k}^{H}(i)-\lambda^{-1}(i)\mathbf{R}^{-1}(i).\label{eq:gvff3}
%\end{split}
%\end{equation}
The adaptive  receiver with the GVFF mechanism \cite{Haykin} is implemented by
using (\ref{eq:filterw})-(\ref{eq:gvff3}) with suitable initial
values.
 By using the error criterion with noise variance weighting and the MSE rather than the instantaneous squared error,
two modified GVFF mechanisms were recently proposed in \cite{ssong3}  and \cite{Leung2}. We refer to them as weighting GVFF (WGVFF) mechanism and
mean squared error GVFF (MGVFF) mechanism, respectively.
Fig. \ref{fig:block} illustrates the block diagram of the
adaptive  receiver with a variable forgetting factor mechanism.
In the following section, we will describe the proposed
low-complexity variable forgetting factor mechanism.}

%\begin{figure}[!hhh]
%\centering \scalebox{0.77}{\includegraphics{MMSEVFFmechanism2.eps}}
%\vspace{-0.95em}\caption{Structure of the adaptive RLS receiver with the variable forgetting factor mechanism } \label{fig:block}
%\end{figure}

\section{Proposed Variable Forgetting Factor Mechanism}
\label{Section4:proposedVFF}

%\subsection{Adaptive TVFF Mechanism}
%
%\begin{equation}
%\lambda(i)=\Big[ \frac{1}{1+\gamma(i)} \Big]^{\lambda^{+}}_{\lambda^{-}}\label{eq:tvff11}
%\end{equation}
%
%\begin{equation}
%\gamma(i)=\delta_{1}\gamma(i-1)+\delta_{2}|b(i)-\mathbf{w}^{H}(i)\mathbf{r}(i)|^2\label{eq:tvff}
%\end{equation}

%In this section, we first introduce the proposed low-complexity
%CTVFF mechanism that adjusts the forgetting factor of the adaptive
%RLS multiuser detector. Then, we present the computational
%complexity analysis for the proposed CTVFF mechanism and the
% existing gradient-based mechanisms.
In this section, we first introduce the proposed low-complexity CTVFF mechanism that adjusts the forgetting factor of the RLS algorithm that equips the adaptive receiver. Then, we present the computational complexity analysis for the proposed CTVFF mechanism and the existing gradient-based mechanisms.

\subsection{Adaptive CTVFF Mechanism}

{  Based on the variable step-size mechanism for least mean square
(LMS) algorithms in \cite{RKwong}, the motivation for the proposed
CTVFF mechanism is that for large {\red estimation errors} the
adaptive algorithm will employ smaller forgetting factors whereas
small {\red estimation errors} will result in an increase of the
forgetting factor in order to yield smaller misadjustment. Thus, the
CTVFF mechanism employs an exponential weighting parameter that
controls the quality of the {\red estimation error}, and
we have devised the following time-averaged
expression}
%\begin{equation}
%\phi(i)=\delta_{1} \phi(i-1)+\delta_{2} (|\mathbf{w}^{H}_{k}(i)\mathbf{r}(i)|^2-1)^2,\label{eq:TAVff2}
%\end{equation}

\begin{equation}
\gamma(i)=\delta_{1}\gamma(i-1)+\delta_{2}\rho^2(i),\label{eq:ctvff1}
\end{equation}
where $0<\delta_{1} <1$ and $\delta_{2}>0$, and $\delta_{1}$ is close to $1$ and $\delta_{2}$ is set to be a small value.
 The quantity $\rho(i)$ denotes the time average estimate related to the correlation of $|b_{k}(i-1)-\mathbf{w}^{H}_{k}(i-1)\mathbf{r}(i-1)|$ and $|b_{k}(i)-\mathbf{w}^{H}_{k}(i)\mathbf{r}(i)|$, which is given by
\begin{equation}
\begin{split}
\rho(i)&=\delta_{3}\rho(i-1)+(1-\delta_{3})|b_{k}(i-1)-\mathbf{w}^{H}_{k}(i-1)\mathbf{r}(i-1)||b_{k}(i)-\mathbf{w}^{H}_{k}(i)\mathbf{r}(i)|,\label{eq:ctvff2}
\end{split}
\end{equation}
where  $0<\delta_{3} <1$, and $\delta_{3}$ controls the time average estimate. We set $\delta_{3}$ to be a quantity close to $1$.

%where $\phi(i)$ denotes an updated component that is controlled by the instantaneous CM cost function,
%%related to the time-averaged CM cost function.
%$0<\delta_{1} <1$, and $\delta_{2}>0$.
The updated component $\rho(i)$ is a small value, and  it changes rapidly as the instantaneous value of the correlation between $|b_{k}(i-1)-\mathbf{w}^{H}_{k}(i-1)\mathbf{r}(i-1)|$ and $|b_{k}(i)-\mathbf{w}^{H}_{k}(i)\mathbf{r}(i)|$.
The use of $\gamma(i)$ has the potential to provide a suitable indication of the evolution of the error correlation.
Thus, unlike the conventional GVFF mechanism we aim to design a
simpler mechanism that adjusts the forgetting factor automatically
based on $\gamma(i)$. Note that the forgetting factor should vary in
an inversely proportional way to the value of the {  error
correlation. We have studied} a number of rules and the following
expression is a result of several attempts to devise a simple and
yet effective mechanism
\begin{equation}
\lambda(i)=\bigg[ \frac{1}{1+\gamma(i)}\bigg]^{\lambda^{+}}_{\lambda^{-}}.\label{eq:ctvff3}
\end{equation}
{  Indeed, the CTVFF mechanism is simple to implement and a detailed
analysis of the algorithm is possible under a few assumptions
commonly made in the literature.} The proposed low-complexity CTVFF
mechanism is given by (\ref{eq:ctvff1}), (\ref{eq:ctvff2}) and
(\ref{eq:ctvff3}).
%
%It is worth to point
%out that other rules have been experimented and the TAVFF scheme is a
%result of several attempts to devise a simple and yet effective
%mechanism.
%
%
The value of variable forgetting factor $\lambda(i)$ is close to $1$, and it is controlled by  the parameters $\delta_{1}$, $\delta_{2}$ and $\delta_{3}$.
We will investigate the sensitivity of the parameters in Section \ref{Section6:simulations}.
%and the %prediction error.
%%In the proposed TAVFF scheme, the forgetting factor  is controlled by the
%instantaneous CM
%cost function.
%Normally, $\delta_{1}$ is close to $1$, and $\delta_{2}$ is set equal to a small value.
%
%The proposed low-complexity variable forgetting factor scheme is given by
%
%\begin{equation}
%\gamma(i)=\bigg[ \frac{1}{1+\gamma^{'}(i)}\bigg]^{\gamma^{+}}_{\gamma^{-}}\label{eq:TAVff1}
%\end{equation}
%where $\gamma^{'}(i)$ denotes an updated component related to the time-averaged CM cost function. %$\Psi^2(i)=(|\mathbf{w}^{H}_{k}(i)\mathbf{r}(i)|^2-1)^2$.
%It uses the following adaptive rule,
%\begin{equation}
%\gamma^{'}(i)=\delta_{1} \gamma^{'}(i-1)+\delta_{2} (|\mathbf{w}^{H}_{k}(i)\mathbf{r}(i)|^2-1)^2\label{eq:TAVff2}
%\end{equation}
%
%where $0<\delta_{1} <1$, and $\delta_{2}>0$. Normally, $\delta_{1}$ is close to $1$, and $\delta_{2}$ is set to be a small value.
%The value of variable forgetting factor $\gamma(i)$ is close to $1$, and it is controlled by  the parameters $\delta_{1}$ and $\delta_{2}$ and the %prediction error.
%%In the proposed TAVFF scheme, the forgetting factor  is controlled by the
%instantaneous CM
%cost function.
A large prediction error will cause the updated component $\gamma(i)$ to increase, which simultaneously reduces the
forgetting factor $\lambda(i)$ and provides a faster tracking. While a small prediction error will result in a decrease in the  updated component $\gamma(i)$, thereby the forgetting factor $\lambda(i)$ is increased to yield a smaller misadjustment.

\subsection{Computational Complexity}

 We describe the computational complexity of the proposed CTVFF and
the adaptive GVFF, WGVFF and MGVFF mechanisms \cite{Haykin, ssong3, Leung2}. Table \ref{tab:table13} shows the
additional computational complexity of the algorithms for multipath
channels.  We estimate the number of arithmetic operations by
considering the number of complex additions and multiplications
required by the mechanisms. The CTVFF and MGVFF mechanisms have constant computational complexity for each received symbol
while  the adaptive GVFF  and  WGVFF  techniques  have  additional
complexity proportional to the length $M$ of the receive filter.
An important advantage of the proposed adaptation rule is that {  it
requires only ten operations.}
%a few fixed number of
%operations.
%while  the adaptive GVFF mechanism \cite{Haykin}  and the WGVFF  mechanism \cite{Leung2} have an additional
%complexity proportional to the length $M$ of the receive filter.
%-------------------------------------
%\begin{table}[h]
%\centering%
%\caption{\normalsize  Additional Computational complexity.} {
%\begin{tabular}{ccc}
%\hline \rule{0cm}{2.5ex}&  \multicolumn{2}{c}{Number of operations
%per  symbol} \\ \cline{2-3}
%Algorithm & {Multiplications} & {Additions} \\
%\hline\\
%%\emph{\small \bf TVFF}  & {\small $$ } & {\small $$ }  \\
%                        % &            &   \\
% \emph{\small \bf CTVFF}  & {\small $7$ } & {\small $3$ }  \\
%                        % &            &   \\
%\emph{\small \bf  GVFF \cite{Haykin}}  &  {\small $7M^2+4M+2$ } & {\small $7M^2+M$ }  \\
%                       % &      &          \\
%\emph{\small \bf  WGVFF \cite{ssong3}}  &  {\small $7M^2+4M+9$ } & {\small $7M^2+M+1$ }  \\
%\emph{\small \bf  MGVFF \cite{Leung2}}  &  {\small $29$ } & {\small $18$ }  \\
%\hline
%\label{tab:table13}
%\end{tabular}
%}
%\end{table}
\begin{table}[h]
\centering%
\caption{\normalsize  Additional Computational complexity.} {
\begin{tabular}{ccc}
\hline \rule{0cm}{2.5ex}&  \multicolumn{2}{c}{Number of operations
per  symbol} \\ \cline{2-3}
Algorithm & {Multiplications} & {Additions} \\
\hline\\
%\emph{\small \bf TVFF}  & {\small $$ } & {\small $$ }  \\
                        % &            &   \\
 \emph{\small \bf CTVFF}  & {\small $7$ } & {\small $3$ }  \\
                        % &            &   \\
\emph{\small \bf  GVFF \cite{Haykin}}  &  {\small $7M^2+4M+2$ } & {\small $7M^2+M$ }  \\
                       % &      &          \\
\emph{\small \bf  WGVFF \cite{ssong3}}  &  {\small $7M^2+4M+9$ } & {\small $7M^2+M+1$ }  \\
\emph{\small \bf  MGVFF \cite{Leung2}}  &  {\small $29$ } & {\small $18$ }  \\
\hline
\label{tab:table13}
\end{tabular}
}
\end{table}

\section{ANALYSES OF THE PROPOSED ALGORITHMS}
\label{Section5:analysis}

%\subsection{Steady-State Analysis for TVFF Mechanism}
%
%\subsection{Steady-State Analysis for CTVFF Mechanism}

%\subsection{Steady-State Mean Values of VFF Mechanisms}

%By using the  expression
%\begin{equation}
%\mathbf{d}_{k}(i)=\gamma(i)\mathbf{d}_{k}(i-1)+z^{*}_{k}(i)\mathbf{r}(i)
%\end{equation}
%and following the same approach we can have (\ref{eq:assum2}).

%\subsection{Convergence Analysis}

%We recall $\mathbf{w}(i)=\mathbf{R}^{-1}(i)\mathbf{z}(i)$ and substitute $b(i)=\mathbf{w}^{H}_{0}\mathbf{r}(i)+\xi_{0}(i)$
%into $\mathbf{z}(i)$, we have
%\begin{equation}
%\begin{split}
%\mathbf{z}(i)&=\sum^{i}_{l=1}\lambda^{i-l}(i)\mathbf{r}(l)\big(\mathbf{r}^{H}(l)\mathbf{w}_{0}+\xi^{*}_{0}(l)\big)
%\\&=\sum^{i}_{l=1}\lambda^{i-l}(i)\mathbf{r}(l)\mathbf{r}^{H}(l)\mathbf{w}_{0}+\sum^{i}_{l=1}\lambda^{i-l}(i)\mathbf{r}(l)\xi^{*}_{0}(l)
%\end{split}
%\end{equation}

In this section, we first show the convergence of the mean weight vector and derive the steady-state MSE expression for the adaptive  receiver with the proposed CTVFF mechanism  in the case of invariant channels. Then, we give the tracking analysis of the proposed scheme for the scenario of time-varying channels. Finally,  the expression for the steady-state mean value of the CTVFF mechanism is derived.

\subsection{Convergence of the Mean Weight Vector}

We show the convergence of the mean weight vector of the proposed algorithm.
Firstly, we give two expressions which will be used for the following derivation:
\begin{equation}
\mathbf{k}(i)=\mathbf{R}^{-1}(i)\mathbf{r}(i)\label{eq:exp1}
\end{equation}
%$\mathbf{k}(i)=\mathbf{R}^{-1}(i)\mathbf{r}(i)$
and
\begin{equation}
\lim_{i\rightarrow \infty}\mathbf{R}^{-1}(i)\approx \big(1-E[\lambda(\infty)]\big)\mathbf{\bar{R}}^{-1}.\label{eq:exp2}
\end{equation}
The proof of (\ref{eq:exp1}) and (\ref{eq:exp2}) are shown in the appendix. The relevant description is given in \cite{tadali}-\cite{macchi}.

Let us recall (\ref{eq:filterw}) and assume
\begin{equation}
b_{k}(i)=\mathbf{w}^{H}_{0}\mathbf{r}(i)+\xi_{0}(i),\label{eq:assu521}
\end{equation}
where $\xi_{0}(i)$ denotes the independent measurement error with zero mean and variance $\sigma^2_{0}$, where $\sigma^2_{0}=1-A_{k}\mathbf{w}^{H}_{0}\mathbf{C}_{k}\mathbf{h}-A_{k}\mathbf{h}^{H}\mathbf{C}^{H}_{k}\mathbf{w}_{0}+\mathbf{w}^{H}_{0}\mathbf{\bar{R}}\mathbf{w}_{0}$, %\Big(\sum^{K}_{k=1}\mathbf{C}_{k}\mathbf{h}\mathbf{h}^{H}\mathbf{C}^{H}_{k}+\sigma^2\mathbf{I}\Big)
which is
computed by (\ref{eq:assu521}).
By defining $\mbox{\boldmath$\epsilon$}(i)=\mathbf{w}_{k}(i)-\mathbf{w}_{0}$ and using (\ref{eq:filterw}) we have
\begin{equation}
\mbox{\boldmath$\epsilon$}(i)=\mathbf{w}_{k}(i)-\mathbf{w}_{0}=\mathbf{w}_{k}(i-1)-\mathbf{w}_{0}+\mathbf{k}(i)e^{*}(i),
\end{equation}
based on (\ref{eq:exp1}) we have
\begin{equation}
\begin{split}
\mbox{\boldmath$\epsilon$}(i)&=\mbox{\boldmath$\epsilon$}(i-1)+\mathbf{R}^{-1}(i)\mathbf{r}(i)\big(b^{*}_{k}(i)-\mathbf{r}^{H}(i)\mathbf{w}_{k}(i-1)\big)
%\\&=\mbox{\boldmath$\epsilon$}(i-1)+\mathbf{R}^{-1}(i)\mathbf{r}(i)\big(\mathbf{r}^{H}(i)\mathbf{w}_{0}+\xi^{*}_{0}(i)-\mathbf{r}^{H}(i)\mathbf{w}_{k}(i-1)\big)
\\&=\mbox{\boldmath$\epsilon$}(i-1)+\mathbf{R}^{-1}(i)\mathbf{r}(i)\big(\xi^{*}_{0}(i)-\mathbf{r}^{H}(i)\mbox{\boldmath$\epsilon$}(i-1)\big)
\\&=\big(\mathbf{I}-\mathbf{R}^{-1}(i)\mathbf{r}(i)\mathbf{r}^{H}(i)\big)\mbox{\boldmath$\epsilon$}(i-1)+\mathbf{R}^{-1}(i)\mathbf{r}(i)\xi^{*}_{0}(i).\label{eq:exp3}
\end{split}
\end{equation}
When $i\rightarrow \infty$, by substituting (\ref{eq:exp2}) into (\ref{eq:exp3}) we obtain
\begin{equation}
\begin{split}
\mbox{\boldmath$\epsilon$}(i)&=\big(\mathbf{I}-\big(1-E[\lambda(\infty)]\big)\mathbf{\bar{R}}^{-1}\mathbf{r}(i)\mathbf{r}^{H}(i)\big)\mbox{\boldmath$\epsilon$}(i-1)+\big(1-E[\lambda(\infty)]\big)\mathbf{\bar{R}}^{-1}\mathbf{r}(i)\xi^{*}_{0}(i).\label{eq:stochasticdifference1}
\end{split}
\end{equation}
{  We note that $1-E[\lambda(\infty)]$ is a small value. According
to the direct averaging method described in \cite{Haykin}, the
solution of the stochastic difference equation
(\ref{eq:stochasticdifference1}), operating under the condition of a
small parameter $1-E[\lambda(\infty)]$, is close to the solution of
another linear stochastic difference equation that is obtained by
replacing the matrix
$(1-E[\lambda(\infty)])\mathbf{r}(i)\mathbf{r}^{H}(i)$ with its
ensemble average $(1-E[\lambda(\infty)])\mathbf{\bar{R}}$. We may
write the new expression as follows}
%
%Since $1-E[\lambda(\infty)]$ is a small value, we use direct averaging \cite{Haykin} and obtain
\begin{equation}
\begin{split}
\mbox{\boldmath$\epsilon$}(i)&\approx\big(\mathbf{I}-\big(1-E[\lambda(\infty)]\big)\mathbf{\bar{R}}^{-1}\mathbf{\bar{R}}\big)\mbox{\boldmath$\epsilon$}(i-1)+\big(1-E[\lambda(\infty)]\big)\mathbf{\bar{R}}^{-1}\mathbf{r}(i)\xi^{*}_{0}(i).
\\&=E[\lambda(\infty)]\mbox{\boldmath$\epsilon$}(i-1)+\big(1-E[\lambda(\infty)]\big)\mathbf{\bar{R}}^{-1}\mathbf{r}(i)\xi^{*}_{0}(i).\label{eq:exp4}
\end{split}
\end{equation}
By taking the expectation of (\ref{eq:exp4}), when $i \rightarrow \infty$, we have
\begin{equation}
E[\mbox{\boldmath$\epsilon$}(i)]\approx E[\lambda(\infty)]E[\mbox{\boldmath$\epsilon$}(i-1)].
\end{equation}
Since $0<E[\lambda(\infty)]<1$, the expected weight error converges
to zero.

%and note that $\mathbf{k}(i)=\mathbf{R}^{-1}(i)\mathbf{r}(i)$, the proof is shown in the appendix.

\subsection{Convergence of MSE}
Then, we show the convergence of MSE for the proposed
algorithm and give an analytical expression to predict the
steady-state MSE.

Based on (\ref{eq:exp4}) we have
\begin{equation}
\begin{split}
\mbox{\boldmath$\Theta$}(i)&=E[\mbox{\boldmath$\epsilon$}(i)\mbox{\boldmath$\epsilon$}^{H}(i)]\\&\approx E^{2}[\lambda(\infty)]E[\mbox{\boldmath$\epsilon$}(i-1)\mbox{\boldmath$\epsilon$}^{H}(i-1)]\\&\quad+\big(1-E[\lambda(\infty)]\big)^2E[\mathbf{\bar{R}}^{-1}\mathbf{r}(i)\xi^{*}_{0}(i)\xi_{0}(i)\mathbf{r}^{H}(i)\mathbf{\bar{R}}^{-1}]
\\&\quad+E[\lambda(\infty)]\big(1-E[\lambda(\infty)]\big)E[\mathbf{\bar{R}}^{-1}\mathbf{r}(i)\xi^{*}_{0}(i)]E[\mbox{\boldmath$\epsilon$}^{H}(i-1)]
\\&\quad+
E[\lambda(\infty)]\big(1-E[\lambda(\infty)]\big)E[\mbox{\boldmath$\epsilon$}(i-1)]E[\xi_{0}(i)\mathbf{r}^{H}(i)\mathbf{\bar{R}}^{-1}].
%\\&=E[\lambda^{2}(\infty)]\mbox{\boldmath$\Theta$}(i-1)+E[\big(1-\lambda(\infty)\big)^2]\mathbf{\bar{R}}^{-1}\sigma^2_{0}
\end{split}
\end{equation}
Note that when $i\rightarrow \infty$, $E[\mbox{\boldmath$\epsilon$}(i-1)]\approx \mathbf{0}$, we have
\begin{equation}
\mbox{\boldmath$\Theta$}(i)\approx E^{2}[\lambda(\infty)]\mbox{\boldmath$\Theta$}(i-1)+\big(1-E[\lambda(\infty)]\big)^2\mathbf{\bar{R}}^{-1}\sigma^2_{0}.
\end{equation}
Since $0<E^{2}[\lambda(\infty)]<1$, we obtain
\begin{equation}
\mbox{\boldmath$\Theta$}(\infty)\approx \Big(\frac{1-E[\lambda(\infty)]}{1+E[\lambda(\infty)]}\Big)\mathbf{\bar{R}}^{-1}\sigma^2_{0}.\label{eq:secorder222}
\end{equation}

Note that the steady-state MSE is given by
%$\lim_{i\rightarrow \infty} \xi(i)=\lim_{i\rightarrow \infty} E[|Ab(i)-\mathbf{w}^{H}_{k}(i-1)\mathbf{r}(i)|^2]\approx
%\lim_{i\rightarrow \infty}\Xi(i)+A^2-\mathbf{w}^{H}_{0}\mathbf{C}_{k}\mathbf{h}-\mathbf{h}^{H}\mathbf{C}^{H}_{k}\mathbf{w}^{H}_{0}$
%-------------------
\begin{equation}
\begin{split}
\lim_{i\rightarrow \infty} \xi(i)&=\lim_{i\rightarrow \infty} E[|b_{k}(i)-\mathbf{w}^{H}_{k}(i-1)\mathbf{r}(i)|^2]\\&=
\lim_{i\rightarrow \infty}(\Xi(i)+1-A_{k}\mathbf{w}^{H}_{0}\mathbf{C}_{k}\mathbf{h}-A_{k}\mathbf{h}^{H}\mathbf{C}^{H}_{k}\mathbf{w}_{0})
%\\&=\lim_{i\rightarrow \infty}\Xi(i)+(1-2\nu)A^2_{k}
,\label{eq:overalMSE1}
\end{split}
\end{equation}
%------------------
% where
% $\mathbf{w}^{H}_{0}\mathbf{C}_{k}\mathbf{h}=\nu$
 and
\begin{equation}
\begin{split}
\Xi(i)
%&=
%E[|\mathbf{w}^{H}_{k}(i-1)\mathbf{r}(i)|^2]\\
&=E[(\mbox{\boldmath$\epsilon$}^{H}(i-1)+\mathbf{w}^{H}_{0})\mathbf{r}(i)\mathbf{r}^{H}(i)(\mbox{\boldmath$\epsilon$}(i-1)
+\mathbf{w}_{0})]
\\&=\mathbf{w}^{H}_{0}\mathbf{\bar{R}}\mathbf{w}_{0}+{\rm
tr}[\mathbf{\bar{R}}\mathbf{\Theta}(i-1)]
+E[\mbox{\boldmath$\epsilon$}^{H}(i-1)]
E[\mathbf{r}(i)\mathbf{r}^{H}(i)\mathbf{w}_{0}]\\
&\quad+E[\mathbf{w}^{H}_{0}\mathbf{r}(i)\mathbf{r}^{H}(i)]E[\mbox{\boldmath$\epsilon$}(i-1)].\label{eq:overalMSE11}
\end{split}
\end{equation}
%
%Since $\lim_{i\rightarrow \infty}E[\mbox{\boldmath$\epsilon$}(i)] =0$,
When $i\rightarrow \infty$, we have $\Xi(\infty)=
\mathbf{w}^{H}_{0}\mathbf{\bar{R}}\mathbf{w}_{0}+{\rm
tr}[\mathbf{\bar{R}}\mathbf{\Theta}(\infty)]$.
%where
%%
%$\Xi_{ex}(i)={\rm tr}[\mathbf{R}\mathbf{\Theta}(i)]$ denotes the steady-state excess MSE.
%
Multiplying (\ref{eq:secorder222}) by $\mathbf{\bar{R}}$ we have
\begin{equation}
{\rm tr}[\mathbf{\bar{R}}\mathbf{\Theta}(\infty)]
\approx\Big(\frac{1-E[\lambda(\infty)]}{1+E[\lambda(\infty)]}\Big)\sigma^2_{0}M.\label{eq:overalMSE2}
\end{equation}
Therefore, the steady-state MSE in (\ref{eq:overalMSE1}) can be
approximated by the following expression
\begin{equation}
\xi(\infty) \approx \xi_{\rm min}
+\Big(\frac{1-E[\lambda(\infty)]}{1+E[\lambda(\infty)]}\Big)\sigma^2_{0}M,\label{eq:ssMSE}
\end{equation}
where $\xi_{\rm min} =1-A^{2}_{k}\mathbf{h}^{H}(i)
\mathbf{C}^{H}_{k}
\mathbf{\bar{R}}^{-1}\mathbf{C}_{k}\mathbf{h}(i)$.

\subsection{Tracking Analysis}

%We examine the convergence properties of the proposed  in a nonstationary environment, for which the optimum
%solution takes on a time-varying form. In this scenario, the blind adaptive algorithm is added a new task to track the minimum point of the error-performance surface, which is no longer fixed.

In a nonstationary scenario, the optimum
solution  takes on a time-varying form. It brings a new
task  to track the  minimum point of the error-performance surface, which is no longer fixed. We investigate the convergence properties of the proposed algorithm in this case.
%--------------------
%We consider the case that the channel varies slowly. The adjacent channel coefficients are assumed to be similar. %$\mathbf{h}(i)\approx\mathbf{h}(i-1)$.
%Thereby,  expression (\ref{eq:filter3}) still holds.
%----------------------------
%thus the pervious conclusions can be reused.
%
In  time-varying channels, the optimum filter coefficients
are considered to vary according to the model $\mathbf{w}_{0}(i)=\mathbf{w}_{0}(i-1)+\mathbf{q}(i)$, where $\mathbf{q}(i)$ denotes a random perturbation \cite{Haykin}. We assume that $\mathbf{q}(i)$ is an  independently generated sequence with zero mean and positive definite autocorrelation matrix $E[\mathbf{q}(i)\mathbf{q}^{H}(i)]$. This
is typical in the context of tracking analyses of adaptive filters \cite{Eweda, widrow}.

By redefining $\mbox{\boldmath$\epsilon$}(i)=\mathbf{w}_{k}(i)-\mathbf{w}_{0}(i)$ and following the aforementioned approach, we have
\begin{equation}
\begin{split}
\mbox{\boldmath$\epsilon$}(i)&=\mathbf{w}_{k}(i)-\mathbf{w}_{0}(i)
\\&=\mathbf{w}_{k}(i-1)-\mathbf{w}_{0}(i-1)-\mathbf{q}(i)+\mathbf{k}(i)e^{*}(i)
%\\&=
%\mbox{\boldmath$\epsilon$}(i-1)+\mathbf{R}^{-1}(i)\mathbf{r}(i)\big(b^{*}(i)-\mathbf{r}^{H}(i)\mathbf{w}(i-1)\big)
\\&=\mbox{\boldmath$\epsilon$}(i-1)-\mathbf{q}(i)+\mathbf{R}^{-1}(i)\mathbf{r}(i)\big(\mathbf{r}^{H}(i)\mathbf{w}_{0}(i-1)+\xi^{*}_{0}(i)-\mathbf{r}^{H}(i)\mathbf{w}_{k}(i-1)\big)
%\\&=\mbox{\boldmath$\epsilon$}(i-1)-\mathbf{q}(i)+\mathbf{R}^{-1}(i)\mathbf{r}(i)\big(\xi^{*}_{0}(i)-\mathbf{r}^{H}(i)\mbox{\boldmath$\epsilon$}(i-1)\big)
\\&=\big(\mathbf{I}-\mathbf{R}^{-1}(i)\mathbf{r}(i)\mathbf{r}^{H}(i)\big)\mbox{\boldmath$\epsilon$}(i-1)+\mathbf{R}^{-1}(i)\mathbf{r}(i)\xi^{*}_{0}(i)-\mathbf{q}(i).
\end{split}
\end{equation}
{ Using (\ref{eq:exp2}) when $i$ becomes large, we obtain
\begin{equation}
\begin{split}
\mbox{\boldmath$\epsilon$}(i)
&\approx (\mathbf{I}-(1-E[\lambda(\infty)])\mathbf{\bar{R}}^{-1}\mathbf{r}(i)\mathbf{r}^{H}(i))\mbox{\boldmath$\epsilon$}(i-1)+\big(1-E[\lambda(\infty)]\big)\mathbf{\bar{R}}^{-1}\mathbf{r}(i)\xi^{*}_{0}(i)-\mathbf{q}(i).\label{eq:stochasticdifference2}
\end{split}
\end{equation}
Since $1-E[\lambda(\infty)]$ is a small value, we may solve (\ref{eq:stochasticdifference2}) for the weight error vector $\mbox{\boldmath$\epsilon$}(i)$ by employing the direct averaging method \cite{Haykin}, which was mentioned previously.
We obtain the following expression}
%Using (\ref{eq:exp2}) and direct averaging \cite{Haykin} we obtain
\begin{equation}
\begin{split}
\mbox{\boldmath$\epsilon$}(i)
&\approx E[\lambda(\infty)]\mbox{\boldmath$\epsilon$}(i-1)+\big(1-E[\lambda(\infty)]\big)\mathbf{\bar{R}}^{-1}\mathbf{r}(i)\xi^{*}_{0}(i)-\mathbf{q}(i).\label{eq:exp45}
\end{split}
\end{equation}
Note that the vector $\mathbf{q}(i)$ is an independent zero mean vector, we have
\begin{equation}
\begin{split}
\mbox{\boldmath$\Theta$}(i)&=E[\mbox{\boldmath$\epsilon$}(i)\mbox{\boldmath$\epsilon$}^{H}(i)]\\&\approx E^{2}[\lambda(\infty)]\mbox{\boldmath$\Theta$}(i-1)+\big(1-E[\lambda(\infty)]\big)^2\mathbf{\bar{R}}^{-1}\sigma^2_{0}+E[\mathbf{q}(i)\mathbf{q}^{H}(i)],\label{eq:exp55}
\end{split}
\end{equation}
and
when $i$ becomes large,
the MSE in a time-varying environment is given by
%$\lim_{i\rightarrow \infty} \xi(i)=\lim_{i\rightarrow \infty} E[|Ab(i)-\mathbf{w}^{H}_{k}(i-1)\mathbf{r}(i)|^2]\approx
%\lim_{i\rightarrow \infty}\Xi(i)+A^2-\mathbf{w}^{H}_{0}\mathbf{C}_{k}\mathbf{h}-\mathbf{h}^{H}\mathbf{C}^{H}_{k}\mathbf{w}^{H}_{0}$
%-------------------
\begin{equation}
\begin{split}
 \xi(i)&= E[|b_{k}(i)-\mathbf{w}^{H}_{k}(i-1)\mathbf{r}(i)|^2]\\&=
\mathbf{w}^{H}_{0}(i-1)\mathbf{\bar{R}}\mathbf{w}_{0}(i-1)+{\rm
tr}[\mathbf{\bar{R}}\mathbf{\Theta}(i)]+1-A_{k}\mathbf{w}^{H}_{0}(i-1)\mathbf{C}_{k}\mathbf{h}-A_{k}\mathbf{h}^{H}\mathbf{C}^{H}_{k}\mathbf{w}_{0}(i-1)
%\\&=\lim_{i\rightarrow \infty}\Xi(i)+(1-2\nu)A^2_{k}
.\label{eq:overalMSE3}
\end{split}
\end{equation}
%------------------
% where
% $\mathbf{w}^{H}_{0}\mathbf{C}_{k}\mathbf{h}=\nu$
%where
%\begin{equation}
%\begin{split}
%\Xi(i)
%%&=
%%E[|\mathbf{w}^{H}_{k}(i-1)\mathbf{r}(i)|^2]\\
%&
%%=E[(\mbox{\boldmath$\epsilon$}^{H}(i-1)+\mathbf{w}^{H}_{0}(i-1))\mathbf{r}(i)\mathbf{r}^{H}(i)(\mbox{\boldmath$\epsilon$}(i-1)
%%+\mathbf{w}_{0}(i-1))]
%=\mathbf{w}^{H}_{0}(i-1)\mathbf{\bar{R}}\mathbf{w}_{0}(i-1)+{\rm tr}[\mathbf{\bar{R}}\mathbf{\Theta}(i-1)]
%%\\&\quad+E[\mbox{\boldmath$\epsilon$}^{H}(i-1)]E[\mathbf{r}(i)\mathbf{r}^{H}(i)\mathbf{w}_{0}(i-1)]\\&\quad+E[\mathbf{w}^{H}_{0}(i-1)\mathbf{r}(i)\mathbf{r}^{H}(i)]E[\mbox{\boldmath$\epsilon$}(i-1)].
%\end{split}
%\end{equation}
By using (\ref{eq:exp55}), when $i$ becomes large, we have
\begin{equation}
{\rm tr}[\mathbf{\bar{R}}\mbox{\boldmath$\Theta$}(i)]\approx
\Big(\frac{1-E[\lambda(\infty)]}{1+E[\lambda(\infty)]}\Big)\sigma^2_{0}M
+\frac{{\rm
tr}\big[\mathbf{\bar{R}}E[\mathbf{q}(i)\mathbf{q}^{H}(i)]\big]}{1-E^2[\lambda(\infty)]}.\label{eq:overalMSE4}
\end{equation}
The MSE for a situation in which the adaptive receiver is tracking a
channel can be computed by the following expression
\begin{equation}
\xi(\infty) \approx \xi_{\rm min}
+\Big(\frac{1-E[\lambda(\infty)]}{1+E[\lambda(\infty)]}\Big)\sigma^2_{0}M
+\frac{{\rm tr}\big[\mathbf{\bar{R}}E[\mathbf{q}(i)
\mathbf{q}^{H}(i)]\big]}{1-E^2[\lambda(\infty)]}.\label{eq:tracMSE}
\end{equation}
Note that we need to compute the quantities $E[\lambda(\infty)]$ and
$E[\mathbf{q}(i) \mathbf{q}^{H}(i)]$ to calculate the above
expression.

\subsection{Steady-State Mean Value of the CTVFF Mechanism}

In order to  derive the expression for the steady-state mean value of the CTVFF mechanism,
we first show the convergence and derive the expression for the steady-state statistical property of the updated component $\gamma(i)$.
%and $\gamma^{'}(i)$.
%namely, $E[\gamma^{'}(\infty)]$ and $E[\gamma^{'2}(\infty)]$.

%Let us recall (\ref{eq:tvff}),
%due to $0<\delta_{1}<1$, by taking the expectation of (\ref{eq:tvff}), we can see that $E[\gamma(i)]$ converges. Note that
%$\lim_{i \rightarrow \infty} E[|b(i)-\mathbf{w}^{H}_{k}(i)\mathbf{r}(i)|^2]=\xi_{min}$, where $\xi_{min}$ denotes the minimum mean square error, $\xi_{min}= 1-\mathbf{h}^{H}\mathbf{C}^{H}_{k}\mathbf{\bar{R}}^{-1}\mathbf{C}_{k}\mathbf{h}$.
%%where $\mathbf{R}=E[\mathbf{r}(i)\mathbf{r}^{H}(i)]=\sum^{K}_{k=1}A^{2}_{k}\mathbf{C}_{k}\mathbf{h}\mathbf{h}^{H}\mathbf{C}^{H}_{k}+\sigma^{2}\mathbf{I}$ and $\xi_{ex}(\infty)$ denotes the steady-state excess error of the
%%CM cost function, $\xi_{min}\gg \xi_{ex}(\infty)$ \cite{hzeng}.
%Subsequently,  we obtain the steady-state mean value of the updated component $\gamma(i)$,
%\begin{equation}
%E[\gamma(\infty)]= \frac{\delta_{2}\xi_{min}}{1-\delta_{1}}.\label{eq:tvffvalue}
%\end{equation}

By recalling (\ref{eq:ctvff2}), the estimate of $\rho(i)$ can be alternatively written as
\begin{equation}
\begin{split}
\rho(i)&=(1-\delta_3)\sum^{i-1}_{l=0}\delta^{l}_{3}|b_{k}(i-l-1)-\mathbf{w}^{H}_{k}(i-l-1)\mathbf{r}(i-l-1)||b_{k}(i-l)-\mathbf{w}^{H}_{k}(i-l)\mathbf{r}(i-l)|\label{eq:rho}
\end{split}
\end{equation}
and by squaring (\ref{eq:rho}) we have
\begin{equation}
\begin{split}
\rho^2(i)&=(1-\delta_3)^2\sum^{i-1}_{l=0}\sum^{i-1}_{j=0}\delta^{l}_{3}\delta^{j}_{3}|b_{k}(i-l-1)-\mathbf{w}^{H}_{k}(i-l-1)\mathbf{r}(i-l-1)|\\&\quad\times|b_{k}(i-j-1)-\mathbf{w}^{H}_{k}(i-j-1)\mathbf{r}(i-j-1)||b_{k}(i-l)-\mathbf{w}^{H}_{k}(i-l)\mathbf{r}(i-l)|\\&\quad\times|b_{k}(i-j)-\mathbf{w}^{H}_{k}(i-j)\mathbf{r}(i-j)|.\label{eq:rho2}
\end{split}
\end{equation}
When $i\rightarrow \infty$, we assume that
$|b_{k}(i-l)-\mathbf{w}^{H}_{k}(i-l)\mathbf{r}(i-l)|$ and
$|b_{k}(i-j)-\mathbf{w}^{H}_{k}(i-j)\mathbf{r}(i-j)|$ are
uncorrelated, thus, we have
$E[|b_{k}(i-l)-\mathbf{w}^{H}_{k}(i-l)\mathbf{r}(i-l)||b_{k}(i-j)-\mathbf{w}^{H}_{k}(i-j)\mathbf{r}(i-j)|]=E[|b_{k}(i-l)-\mathbf{w}^{H}_{k}(i-l)\mathbf{r}(i-l)|]E[|b_{k}(i-j)-\mathbf{w}^{H}_{k}(i-j)\mathbf{r}(i-j)|]\approx
0$, where $\forall j\neq l$. Hence, the expectation of $\rho^2 (i)$
is given by
\begin{equation}
\begin{split}
E[\rho^2 (i)]&=(1-\delta_3)^2\sum^{i-1}_{l=0}\delta^{2l}_{3} E[|b_{k}(i-l-1)-\mathbf{w}^{H}_{k}(i-l-1)\mathbf{r}(i-l-1)|^2] E[|b_{k}(i-l)-\mathbf{w}^{H}_{k}(i-l)\mathbf{r}(i-l)|^2].
\end{split}
\end{equation}
Note that $0<\delta_{3}<1$, by using $\lim_{i \rightarrow \infty} E[|b_{k}(i)-\mathbf{w}^{H}_{k}(i)\mathbf{r}(i)|^2]=\xi_{min}$, we obtain
\begin{equation}
E[\rho^2 (i)]=\frac{(1-\delta_{3})\xi^2_{min}}{(1+\delta_{3})}.
\end{equation}
Note that $0<\delta_{1}<1$, based on (\ref{eq:ctvff1}) we obtain the steady-state mean  for $\gamma(i)$,
\begin{equation}
E[\gamma(\infty)]= \frac{\delta_{2}(1-\delta_{3})\xi^2_{min}}{(1-\delta_{1})(1+\delta_{3})}.\label{eq:ctvffvalue}
\end{equation}
From  (\ref{eq:ctvff1}) and (\ref{eq:ctvff2}), we can see that the
quantity $\gamma(i)$  is a  small value, and $\gamma(i)$ varies
slowly around its mean value. Thus, when $i\rightarrow \infty$, by
using (\ref{eq:ctvff3}) and (\ref{eq:ctvffvalue}) we assume that
the  steady-state mean value of the variable forgetting factor for
the CTVFF mechanism is given as
\begin{equation}
\begin{split}
%E[\alpha(\infty)]=\frac{1-\delta_{1}}{1+\delta_{2}\xi_{min}-\delta_{1}}\label{eq:vff1}
E[\lambda(\infty)]&\approx\frac{1}{1+E[\gamma(\infty)]}\\
&=\frac{(1-\delta_{1})(1+\delta_{3})}{(1-\delta_{1})(1+\delta_{3})+\delta_{2}(1-\delta_{3})\xi^{2}_{min}}.\label{eq:vff1}
\end{split}
\end{equation}
The above value can be substituted in (\ref{eq:ssMSE}) and
(\ref{eq:tracMSE}) to predict the MSE.

\section{Simulations}
\label{Section6:simulations}

In this section, we first adopt a simulation approach and conduct
several experiments in order to verify the effectiveness of the
proposed CTVFF mechanism in RLS algorithms applied to interference
suppression problems with DS-CDMA systems. Then, we verify the
effectiveness of the proposed analytical expressions in
(\ref{eq:ssMSE}), (\ref{eq:tracMSE}) and (\ref{eq:vff1}) to predict
the performance of the adaptive linear receivers in several
situations. The DS-CDMA system employs Gold sequences as the
spreading codes, and the spreading gain is $N=15$. The sequence of
channel coefficients for each path is given by
$h_{f}(i)=p_{f}\alpha_{f}(i)(f=0,1,2)$. All channels are normalized
so that
\begin{equation}
\sum^{L_{p}-1}_{f=0}p^{2}_{f}=1,
\end{equation}
%have a
%profile with three paths whose powers are $p_{0}=0$ dB,
%$p_{1}=-7$ dB and $p_{2}=-10$ dB, respectively,
%
where
$\alpha_{f}(i)$ is computed according to the Jakes' model \cite{rappaport}.
%We assume that the  MIMO broadcast channel is quasi-static flat fading with
%Rayleigh distribution.
%The channel varies per transmit block, each block contains $500$ symbols, and
%$1000$ channel realizations are employed for each simulation.
%We consider packets with $1000$ symbols.
For the initial values, we set $\mathbf{R}^{-1}(0)=\mathbf{I}$, $\mathbf{w}_{k}(0)=0.01\times\mathbf{1}$, $\frac{\mathbf{\partial \mathbf{w}_{k}(0)}}{\partial \lambda}=\mathbf{0}$ and
$\frac{\mathbf{\partial \mathbf{R}^{-1}(0)}}{\partial \lambda}=\mathbf{I}$, where $\mathbf{1}$ denotes an all-one vector.  The parameters of the MGVFF and WGVFF mechanisms are set as \cite{Leung2, ssong3}.
The simulations are averaged over $10000$ runs. We set the power of the desired user $|A_{1}|^2=1$.

\subsection{Effects of $\delta_{1}$, $\delta_{2}$ and $\delta_{3}$}

\begin{figure}[!hhh]
\centering \scalebox{0.48}{\includegraphics{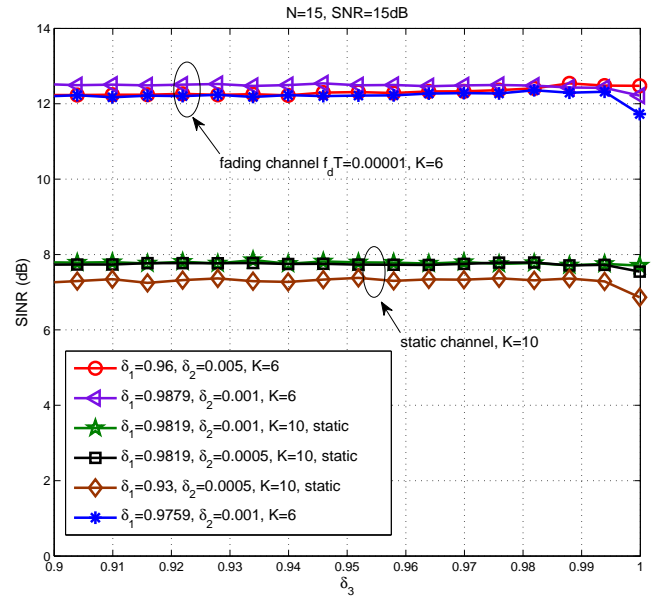}}
\caption{Steady-state SINR versus $\delta_{3}$ for different values
of $\delta_{1}$ and $\delta_{2}$. $SNR=15$ dB. }
\label{fig:simulation1}
\end{figure}

%We first investigate the effects of
Fig. \ref{fig:simulation1} shows the effect of $\delta_{3}$ on the proposed CTVFF mechanism. We show the received steady-state signal to interference plus noise ratio (SINR) versus
$\delta_{3}$ for different values of $\delta_{1}$ and $\delta_{2}$ under different scenarios. For the first scenario, we consider that the system has six users including two users operating at a power level $3dB$ above and one user operating at a power level $6dB$ above the desired user's power level. The channel fading rate is $f_{d}T=1\times 10^{-5}$. The channel power profile is given by  $p_{0}=0 dB$, $p_{1}=-6 dB$ and $p_{2}=-10 dB$. For the second scenario, the system has ten users which have the same power level. In this case, we consider a static channel, the channel parameters of which are given by $h_{0}=0 dB$, $h_{1}=-3 dB$ and $h_{2}=-6 dB$. In  Fig. \ref{fig:simulation1}, we can see that the performance does not change too much as the value of $\delta_{3}$ varies for different values of $\delta_{1}$ and $\delta_{2}$.  We set $\gamma^{-}=0.98$ and $\gamma^{+}=0.99998$, and
use $15 dB$ for the input signal to noise ratio (SNR).
We tuned $\delta_{3}=0.99$ for the following simulations.

\begin{figure}[!hhh]
\centering \scalebox{0.42}{\includegraphics{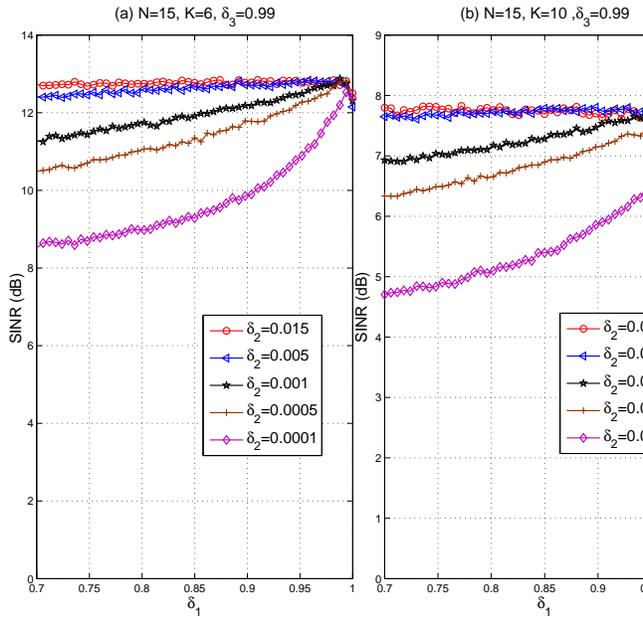}} \caption{
Steady-state SINR versus $\delta_{1}$ for different values of
$\delta_{2}$.  (a) $K=6$, with equal power level interferers, (b)
$K=10$, with equal power level interferers.  $\delta_{3}=0.99$,
static channel.} \label{fig:simulation2}
\end{figure}

We investigate the effects of  $\delta_1$ and $\delta_2$ on the CTVFF mechanism, we show  the received steady-state SINR versus $\delta_1$ for $\delta_2=0.015, 0.005, 0.001, 0.0005, 0.0001$ under different scenarios, where we tuned $\delta_{3}=0.99$. We use $15 dB$ for the input SNR, and set $\gamma^{-}=0.98$ and $\gamma^{+}=0.99998$ for the CTVFF mechanism to guarantee the
stability. The results shown in Fig. \ref{fig:simulation2} (a) and (b) are based on a static state channel.
The time-invariant channel parameters are given by $h_{0}=0 dB$, $h_{1}=-3 dB$ and $h_{2}=-6 dB$.
In Fig. \ref{fig:simulation2} (a), the system has six users with the same power level.
% %including three
%users operating at a power level $3 dB$ above
% and one user operating at a power level $6 dB$  above the desired user's power level.
 In Fig. \ref{fig:simulation2} (b), the system has ten users with the same power level.  It is observed that, firstly, the optimum choice of the pair  $(\delta_1, \delta_2)$ is not unique, for each value of $\delta_2$ we can find a relevant $\delta_1$ to have the best performance. Secondly, with the increasing of $\delta_2$ the performance becomes less sensitive to $\delta_1$. Moreover, when $\delta_{2}$ increases to $0.005$, the performance becomes insensitive to $\delta_1$.
%Fourthly, $( 0.9822, 0.00015)$ is one of the best choice for $\delta_1$ and $\delta_2$ for both scenarios.
From the results, we can obtain that $\delta_1=0.9819$ and $\delta_2=0.005$ are one of the best choices for the scenario in Fig. \ref{fig:simulation2} (a) and
$\delta_1=0.9879$ and $\delta_2=0.001$ are one of the best choices for  both scenarios, which shows that it is possible for us to find a pair $(\delta_{1}, \delta_{2})$ that works well for different scenarios.

%$\delta_1=0.9822$ and $\delta_2=0.00015$ are  the one of optimum selections for both scenarios.

\begin{figure}[!hhh]
\centering \scalebox{0.42}{\includegraphics{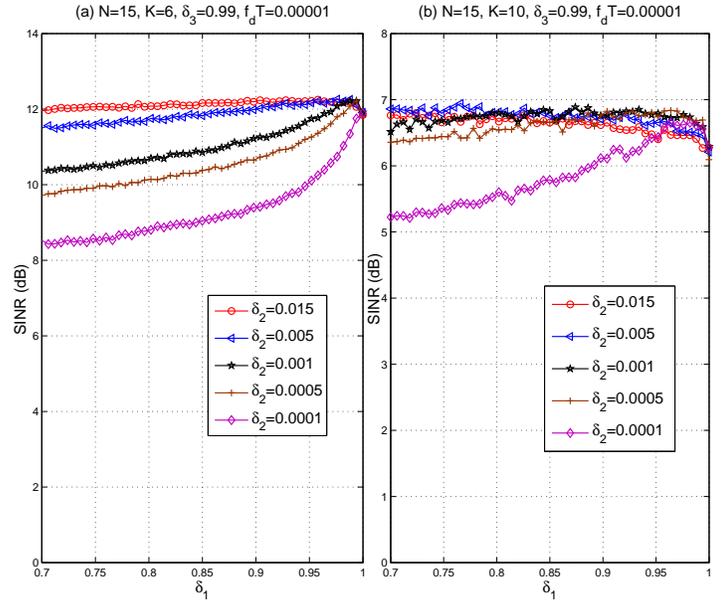}} \caption{
Steady-state SINR versus $\delta_{1}$ for different values of
$\delta_{2}$.  (a) $K=6$, with two $3$dB and one $6$dB high power
level interferers, (b) $K=10$, with four $3$dB and two $6$dB high
power level interferers. $f_{d}T=0.00001$, $\delta_{3}=0.99$.}
\label{fig:simulation3}
\end{figure}

The results  in Fig. \ref{fig:simulation3} (a) and (b) show the steady-state SINR versus $\delta_1$ for different values of $\delta_2$ based on the channel with  $f_dT=0.00001$.
The channel has a power profile given by $p_{0}=0 dB$, $p_{1}=-6 dB$ and $p_{2}=-10 dB$.
We tuned $\delta_{3}=0.99$.
In Fig. \ref{fig:simulation3} (a), the system includes six users including two
users operating at a power level $3 dB$ above
and one user operating at a power level $6 dB$  above the desired user's power level. In Fig. \ref{fig:simulation3} (b), the system has ten users including four users operating at a power level $3 dB$ above and two users operating at a power level $6 dB$ above the desired user's power level.
%We tuned $\delta_{3}=0.99$.
We can see that the previous findings on effects of $\delta_1$ and $\delta_2$ still hold for the experiments.
In particular, in Fig. \ref{fig:simulation3} (a), when $\delta_{2}$ increases to $0.015$, the performance becomes insensitive to $\delta_{1}$.
In this case, we found that $\delta_1= 0.934$ and $\delta_2= 0.005$ are one of the best choices for both scenarios.
In the following experiments, for a given scenario we first choose the optimum parameters, and then fix them through the simulations.
In practice, these optimized values should be obtained by experimentation for a given range of values or a specific operating point, and stored at the receiver.

 %In Fig. \ref{fig:simulation3},
%%we study the convergence behavior for different values of $\delta_1$ and  $\delta_2$. W
%w%e show the SINR performance versus the number of received symbols in terms of the  simulation environment of Fig. \ref{fig:simulation3} (a), where
%t%hree pairs $(0.9970, 0.000015)$, $(0.9822,0.00015)$ and $(0.9896,0.00008)$ are employed for $\delta_1$ and $\delta_2$.
 %, where $(0.9896,0.00008)$ is the best selection obtained from Fig. \ref{fig:simulation1} (a).
% From the results in Fig.  \ref{fig:simulation3}, we can see that the three choices provide almost the same convergence performance, which shows that
%it is possible for us to find a pair $(\delta_1, \delta_2)$ that works well for different scenarios.
%%It shows that there exists a pair $(\delta_1, \delta_2)$ that works well for different scenarios.
%In the following experiments, for a given scenario we first choose the optimum parameters, and then fix them through the simulations.
%In practice, these optimized values should be obtained by experimentation for a given range of values or a specific operating point,
%and stored at the receiver.

\subsection{Performance of the Proposed Algorithm}

\begin{figure}[!hhh]
\centering \scalebox{0.53}{\includegraphics{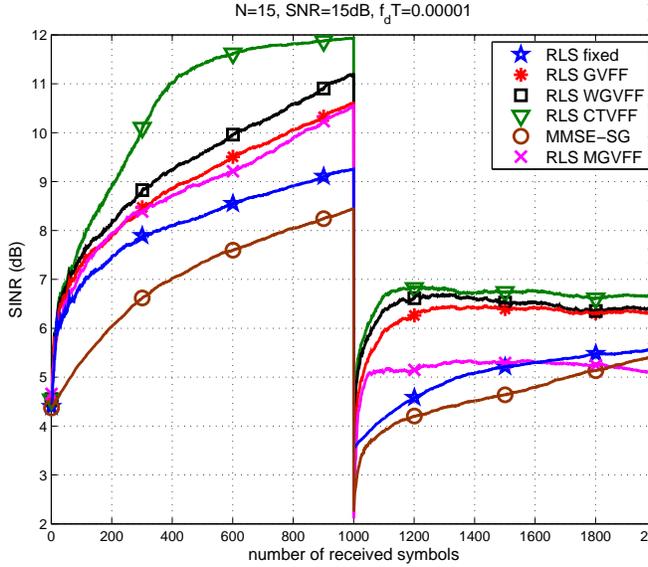}} \caption{
SINR performance in nonstationary environment of multipath fading
channels.  SNR$=15$dB. $f_{d}T=0.00001$.} \label{fig:simulation4}
\end{figure}

We first choose the received SINR as the performance index to evaluate the convergence performance in nonstationary scenarios.
In this simulation, we assess the SINR performance of  the proposed CTVFF mechanism, the adaptive GVFF mechanism,
 the MGVFF mechanism, the WGVFF mechanism,
the fixed forgetting factor mechanism and the adaptive SG receiver.
The results shown in Fig. \ref{fig:simulation4} illustrate that the performance
in terms of SINR of the analyzed algorithms in a nonstationary scenario.
The channel has a power profile given by $p_{0}=0 dB$, $p_{1}=-6 dB$ and $p_{2}=-10 dB$.
The
system starts with six users including two users operating at a power level $3$ dB above and one user operating at a power level
$6$ dB above the desired user's power level.
At $1000$ symbols, four interferers including two users operating at a power level $3$ dB above and one user operating at a power level $6$ dB above the desired user's power level  enter the system.
The normalized Doppler frequency is $f_{d}T=1\times 10^{-5}$. From Fig. \ref{fig:simulation4}, we can see that the proposed CTVFF mechanism with the adaptive  receiver achieves the best performance, followed by
 the WGVFF mechanism with the RLS receiver and
the  GVFF mechanism with the  RLS receiver.
%the MGVFF mechanism with the RLS receiver,
%the adaptive RLS receiver  with the fixed forgetting factor mechanism and the adaptive SG receiver.
The MGVFF mechanism does not work well in the nonstationary scenario of multipath fading channels.
The algorithms process $250$
symbols in the TR mode.
%For the simulation, we tuned the parameters of the
%mechanisms, as shown in Table \ref{tab:table2}.
%\begin{table}[h]
%\centering%
%\caption{\small  OPTIMIZED PARAMETERS FOR CASE $2$} {
%\begin{tabular}{cc}
%%\hline \rule{0cm}{2.5ex}&
%% \multicolumn{2}{c}{Number of operations
%%%per  symbol} \\
%% \hline
%%Algorithm & Multiplications & Additions \\
%\hline
%\emph{\small \bf Fixed Schemes}  & {$\gamma=0.9992$ }  \\
%\hline
%                        % &            &   \\
%\emph{\small \bf Blind GVFF}  & {$\gamma(0)=0.999$, $\mu=0.0001$, $\frac{\partial \mathbf{Q}^{-1}_{k}(0)}{\partial \gamma}=\mathbf{I}$, }\\  & {$\mathbf{Y}_{k}(0)=0.01\times\mathbf{1}$, $\frac{\partial \mathbf{d}_{k}(0)}{\partial \gamma}=\mathbf{0}$, $\mathbf{s}_{k}(0)=\mathbf{0}$ }  \\
%& $\gamma^{-}=0.98$, $\gamma^{+}=0.99998$  \\
%                       % &      &          \\
%\hline
%\emph{\small \bf TAVFF}  &   { $\phi(0)=0$, $\delta_{1}=0.9970$, $\delta_{2}=0.000015$ }  \\
%& $\gamma^{-}=0.95$, $\gamma^{+}=0.99998$  \\
%\hline
%\label{tab:table2}
%\end{tabular}
%}
%\end{table}
For the simulation, we tuned the forgetting factor $\lambda=0.997$ for the fixed forgetting factor scheme.
For the GVFF mechanism, we tuned $\lambda^{-}=0.992$, $\lambda^{+}=0.99998$, $\lambda(0)=0.998$ and $\mu=0.0025$.
For the proposed CTVFF mechanism, we tuned $\lambda^{-}=0.98$, $\lambda^{+}=0.99998$, $\delta_{1}=0.934$, $\delta_{2}=0.005$, $\delta_{3}=0.99$, $\gamma(0)=0$ and
$\rho(0)=0$.

\begin{figure}[!hhh]
\centering \scalebox{0.5}{\includegraphics{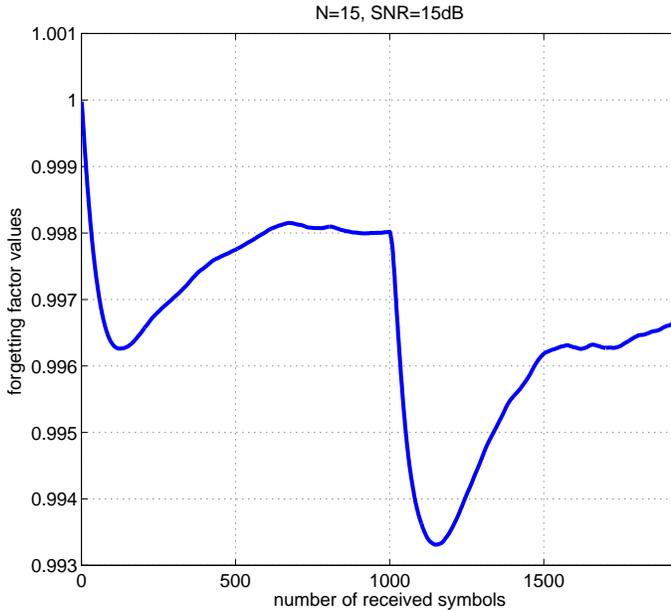}}
\caption{Forgetting factor variation of the proposed CTVFF scheme in
nonstationary scenarios. SNR$=15$dB, $f_{d}T=1\times10^{-5}$.}
\label{fig:simulation5}
\end{figure}

The result shown in Fig. \ref{fig:simulation5} illustrates  the
variation of the forgetting factor value of the  proposed CTVFF
mechanism versus number of received symbols in the  experiment of Fig. \ref{fig:simulation4}.
%The curve corresponds to the  proposed CTVFF mechanism.
We can see that the forgetting factor value becomes small at the first several iterations due to the large prediction error and it increases gradually until the system converges.
At $1000$ symbols, when the new interferers enter the system, the forgetting factor will be recomputed.
%Because the second experiment introduces interferers with higher power,
%the forgetting factor value of the first experiment is greater than the one of the second experiment.

\begin{figure}[!hhh]
\centering \scalebox{0.53}{\includegraphics{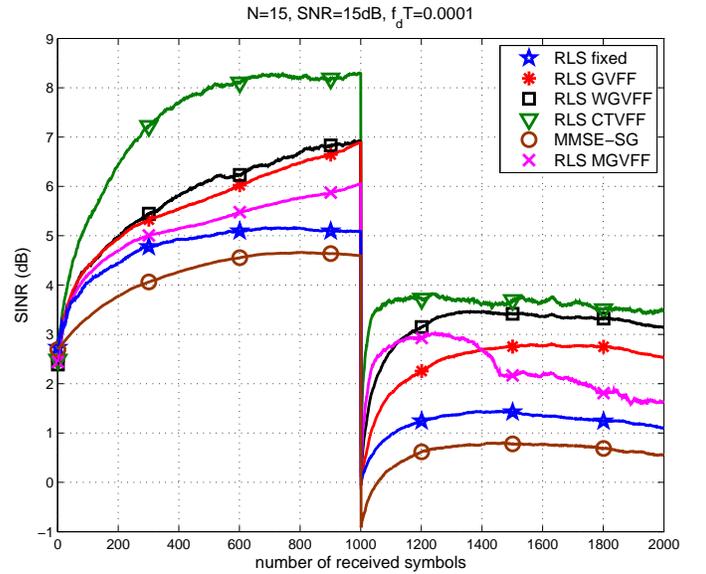}} \caption{ {
SINR performance in nonstationary environment of multipath fading
channels.  SNR$=15$dB. $f_{d}T=0.0001$.}} \label{fig:simulation52}
\end{figure}

{ In Fig. \ref{fig:simulation52}, we depict the SINR performance of
the analyzed algorithms in a different nonstationary scenario. The
channel has a power profile given by $p_{0}=0dB$, $p_{1}=-3dB$ and
$p_{2}=-6dB$. The normalized Doppler frequency is
$f_{d}T=1\times10^{-4}$. The system starts with eight users
including three users operating at a power level $3 dB$ above and
one user operating at a power level $6 dB$ above the desired user's
power level. At $1000$ symbols, four interferers including two users
operating at a power level $3 dB$ above the desired user's power
level enter the system. From the results, we can see that the
adaptive receiver with the  CTVFF mechanism works well and still
outperforms the conventional techniques. For the proposed CTVFF
mechanism,
%we tuned $\lambda^{-}=0.98$, $\lambda^{+}=0.99998$, $\delta_{1}=0.934$, $\delta_{2}=0.005$ and $\delta_{3}=0.99$.
we maintained the parameters in the simulation of Fig. \ref{fig:simulation4}.
%$\gamma(0)=0$ and $\rho(0)=0$.
We tuned the forgetting factor $\lambda=0.995$ for the fixed forgetting factor scheme. For the GVFF mechanism, we tuned $\lambda^{-}=0.992$, $\lambda^{+}=0.99998$ and $\mu=0.0035$.
}

\begin{figure}[!hhh]
\centering \scalebox{0.53}{\includegraphics{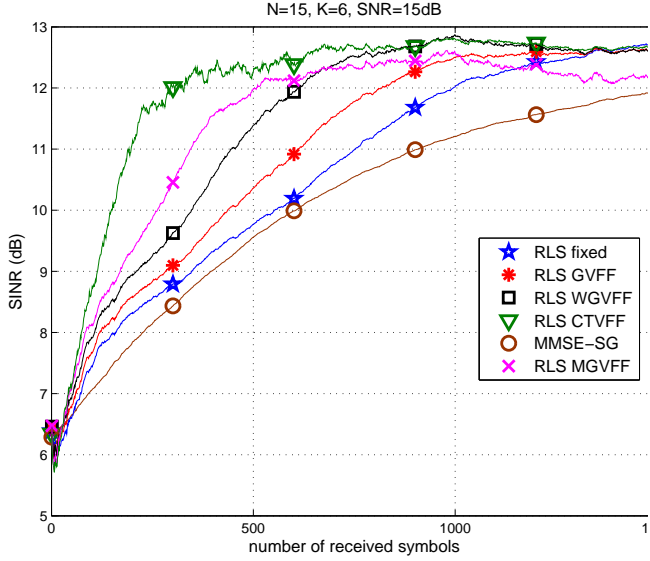}} \caption{
SINR performance in a static environment.  SNR$=15$dB. $K=6$.}
\label{fig:simulation6}
\end{figure}

We examine the SINR performance of the CTVFF mechanism in a static environment. We compare the CTVFF mechanism with the  adaptive GVFF mechanism,
the MGVFF mechanism, the WGVFF mechanism, the fixed forgetting factor mechanism and the adaptive SG receiver. Fig. \ref{fig:simulation6}
illustrates the SINR performance versus the number of received symbols for a static environment, where the time-invariant channel parameters are given by $h_{0}=0dB$, $h_{1}=-3dB$ and $h_{2}=-6dB$. The system includes six users which have equal power level. The results show that
the proposed CTVFF mechanism works well in a static environment, and it outperforms the other conventional schemes. For the simulation, we tuned the forgetting factor $\lambda=0.9995$ for the fixed forgetting factor scheme.
For the GVFF mechanism, we tuned $\lambda^{-}=0.992$, $\lambda^{+}=0.99998$, $\lambda(0)=0.998$ and $\mu=0.006$.
For the proposed CTVFF mechanism, we tuned $\lambda^{-}=0.98$, $\lambda^{+}=0.99998$, $\delta_{1}=0.9879$, $\delta_{2}=0.001$ and $\delta_{3}=0.99$.

%\begin{figure}[!hhh]
%\centering \scalebox{0.53}{\includegraphics{analysis1_new.eps}}
%\vspace{-1.5em}\caption{\small Analytical MSE versus simulated performance for convergence and tracking
%analyses of the proposed CTVFF scheme. $250$ symbols in TR.}
%\label{fig:simulation5}
%\end{figure}
%
%\begin{figure}[!hhh]
%\centering \scalebox{0.53}{\includegraphics{analysis2_new.eps}}
%\vspace{-1.5em}\caption{\small Analytical steady-state MSE versus simulated performance for the proposed CTVFF scheme in time-invariant and fading channels. $1500$ symbols are transmitted, $250$ symbols in TR.}
%\label{fig:simulation6}
%\end{figure}

%\subsection{BER Performance}

\begin{figure}[!hhh]
\centering \scalebox{0.48}{\includegraphics{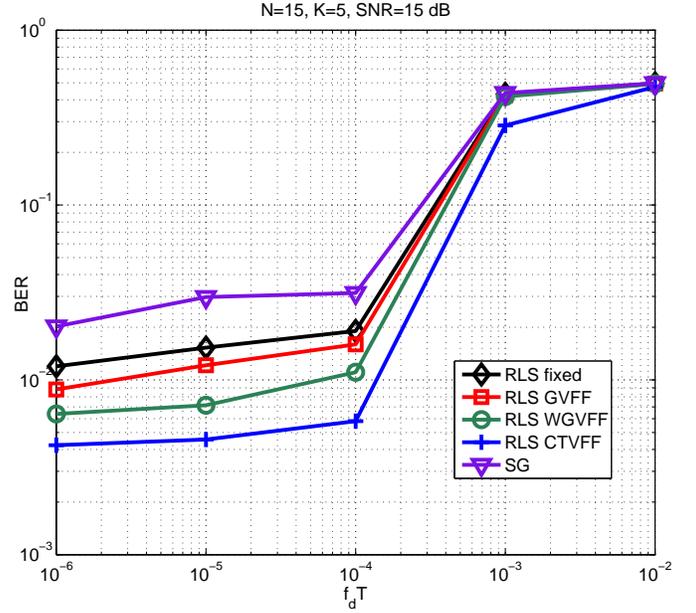}}
\vspace{-1.5em}\caption{{ \small BER performance versus the
(cycles/symbol) in multipath fading channels. We use SNR$=15$dB,
K$=5$, $1500$ symbols are transmitted and $250$ symbols in TR.}}
\label{fig:fadingrate}
\end{figure}

We show the bit error rate (BER) performance of the following
algorithms as the fading rate of the channel varies, i.e., the  RLS
receiver with the proposed CTVFF mechanism, the RLS receiver with
the GVFF mechanism,
 the RLS receiver with the WGVFF mechanism,
the RLS receiver with the fixed forgetting
factor mechanism and the adaptive SG receiver. In this experiment,
the values of the forgetting factors for the algorithms which employ
fixed forgetting factors are optimized, and we tuned suitable
forgetting factors for each value of $f_{d}T$.
The channel has a  power profile given by $p_{0}=0dB$, $p_{1}=-6dB$ and $p_{2}=-10dB$.
In Fig.
\ref{fig:fadingrate}, we can see that, as the fading rate increases,
the performance gets worse, and our proposed CTVFF scheme
outperforms the existing schemes, which shows the ability of the RLS
receiver with the CTVFF mechanism to deal with  channel
uncertainties. For the GVFF mechanism, we tuned $\lambda^{-}=0.993$, $\lambda^{+}=0.99998$ and $\mu=0.003$.
For the proposed CTVFF mechanism, we tuned $\lambda^{-}=0.98$, $\lambda^{+}=0.99998$, $\delta_{1}=0.988$, $\delta_{2}=0.001$ and $\delta_{3}=0.99$.

\begin{figure}[!hhh]
\centering \scalebox{0.44}{\includegraphics{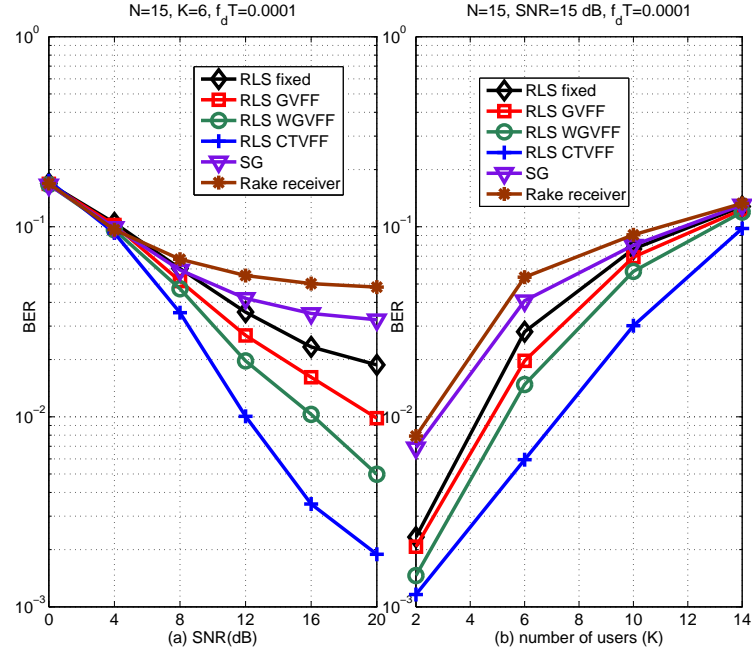}} \caption{
SINR performance versus (a) SNR and (b) the number of users (K) in
multipath time varying channels, $f_{d}T=0.0001$.}
\label{fig:simulation8}
\end{figure}

Fig. \ref{fig:simulation8} (a) and (b) illustrate
the  BER performance of the desired user versus SNR and number of
users $K$, respectively, where we set  $f_{d}T=1\times10^{-4}$.  We can see that
the best performance is achieved by the  adaptive  receiver with
the CTVFF mechanism, followed by the RLS receiver with the WGVFF
mechanism,
 the RLS receiver with the GVFF mechanism,
the RLS receiver with the fixed forgetting factor
mechanism, the  adaptive SG receiver and the conventional Rake
receiver. In particular, the adaptive  receiver with the CTVFF
mechanism   can save up to over $4$dB and support up to two more
users in comparison with the   receiver with the WGVFF
 mechanism, at the BER level of $ 10 ^{-2}$.
% The
%algorithms process $250$ symbols in TR and $1500$ symbols in DD.
$1500$ symbols are transmitted and $250$ symbols are used in TR.

\subsection{Convergence and Tracking Analyses}

\begin{figure}[!hhh]
\centering \scalebox{0.5}{\includegraphics{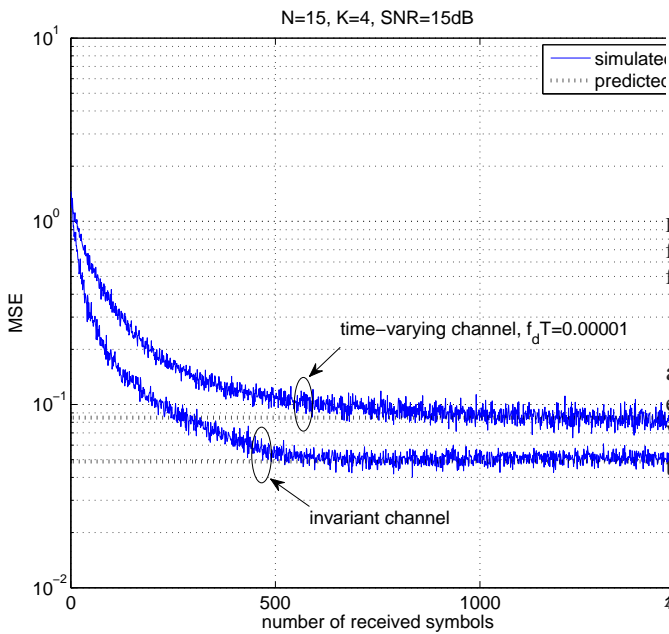}}
\vspace{-1.5em}\caption{\small Analytical MSE versus simulated
performance for convergence and tracking analyses of the proposed
CTVFF scheme. $250$ symbols in TR.} \label{fig:simulation9}
\end{figure}

Then, we examine the convergence and tracking analyses of the
proposed CTVFF mechanism with the RLS receiver. The steady-state MSE
between the desired and the estimated symbol obtained through
simulation is compared with the steady-state MSE computed via the
expressions derived in Section V. We verify  the analytical results
(\ref{eq:ssMSE}) and (\ref{eq:vff1}) to predict the
steady-state MSE in time-invariant channels and the analytical results
(\ref{eq:tracMSE}) and (\ref{eq:vff1}) to predict the
steady-state MSE in time-varying fading channels. In this simulation, we assume
that four users having the same power level operate in the system.
The channel has a power profile given by $p_{0}=0dB$, $p_{1}=-6dB$ and $p_{2}=-10dB$.
% All experiments are averaged over $1000$
%runs.
We employ $\delta_{1}=0.99$, $\delta_{2}=0.35\times 10^{-2}$ and
$\delta_{3}=0.995$ for the case of invariant channels, and employ
$\delta_{1}=0.99$, $\delta_{2}=0.4\times 10^{-3}$ and
$\delta_{3}=0.99$ for the case of time-varying fading channels, where
$f_{d}T=1\times10^{-5}$. $250$ symbols are used in TR.
%We consider the case that the channel varies
%slowly.
% %The variance of the element in $\mathbf{q}(i)$, i.e., $\frac{{\rm tr}\big[E[\mathbf{q}(i)\mathbf{q}^{H}(i)] \big]}{M}$ is
%c%omputed with the aid of $J_{0}(2\pi f_{d}T)$ \cite{besselj1}, where $J_{0}$ is the
%z%ero-order Bessel function of the first kind. The quantity of $E[\mathbf{q}(i)\mathbf{q}^{H}(i)]$ is estimated by $\big(\sum^{N_{e}}_{i=1}\mathbf{q}(i)\mathbf{q}^{H}(i)\big)/N_{e}$, where %$%N_{e}$ is set to be a large number.
The quantity of $E[\mathbf{q}(i)\mathbf{q}^{H}(i)]$ is estimated by using the average over  $10000$ independent experiments, namely,
$\big(\sum^{N_{e}}_{i=1}\mathbf{q}(i)\mathbf{q}^{H}(i)\big)/N_{e}$, where $N_{e}=10000$ and $\mathbf{q}(i)=\mathbf{w}_{0}(i)-\mathbf{w}_{0}(i-1)$.
By comparing the curves  in
Fig. \ref{fig:simulation9}, it can
be seen that as the number of received symbols increases and the
simulated MSE values converge to the analytical results, showing the
usefulness of our analyses and assumptions.
%Fig. \ref{fig:simulation6} (b) shows the
%%effect that the desired
%%user's signal-to-noise power ratio has on the MSE,
%MSE performance versus the desired user's SNR,
% and a comparison
%between the steady-state analysis and simulation results.
%The simulation and analysis results agree well
%with each other.

\begin{figure}[!hhh]
\centering \scalebox{0.5}{\includegraphics{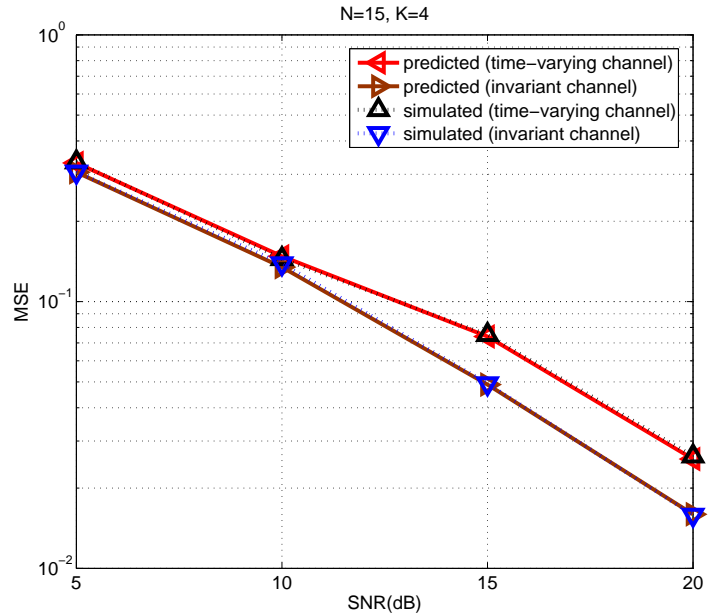}}
\vspace{-1.5em}\caption{\small Analytical steady-state MSE versus
simulated performance for the proposed CTVFF scheme in
time-invariant and time-varying fading channels. $1500$ symbols are
transmitted, $250$ symbols in TR.} \label{fig:simulation10}
\end{figure}

Fig. \ref{fig:simulation10}  shows the
effect that the desired user's SNR has on the MSE, and a comparison between the steady-state analysis and simulation results. We assume that four users operate with the same  power level in the system. We can see that the simulation and analysis results agree well
with each other.
% Table \ref{tab:table2} and \ref{tab:table3} give the simulation parameters of the time-invariant and time-varying fading channels in the simulation, respectively.
  $1500$ symbols are transmitted and $250$ symbols are used in TR. For the time-varying fading channel, we use $f_{d}T=1\times10^{-5}$.

\section{Conclusion}
\label{Section7:conclusion}

 In this paper, we proposed a low-complexity variable
forgetting factor mechanism for
estimating the parameters of CDMA multiuser detector that operate
with RLS algorithms.
 We compared
the computational complexity of the new algorithm with the
existing gradient-based methods and further investigated
 the convergence and tracking analyses of the proposed CTVFF scheme.
 We also derived expressions to predict the steady-state MSE of the adaptive RLS algorithm with the CTVFF mechanism.
%The CSI quantization scheme base on Lloyd algorithm
%and three methods for interleaving codebook design were described.
The simulation results verify the analytical results and show that the proposed scheme significantly outperforms existing algorithms and supports
systems with higher loads. We remark that our proposed algorithm
also can be extended to take into account
%beamforming, speech processing and
other types of applications.

\begin{appendix}

\subsection{Proof of (\ref{eq:exp1})}

By multiplying $\mathbf{r}(i)$ on  both sides of (\ref{eq:filter0}) we have
\begin{equation}
\begin{split}
\mathbf{R}^{-1}(i)\mathbf{r}(i)&=\lambda^{-1}(i)\mathbf{R}^{-1}(i-1)\mathbf{r}(i)-\lambda^{-1}(i)\mathbf{k}(i)\mathbf{r}^{H}(i)\mathbf{R}^{-1}(i-1)\mathbf{r}(i).\label{eq:filter1}
\end{split}
\end{equation}
Rewrite (\ref{eq:k}) as
\begin{equation}
\begin{split}
\mathbf{k}(i)&=\lambda^{-1}(i)\mathbf{R}^{-1}_{k}(i-1)\mathbf{r}(i)-\lambda^{-1}(i)\mathbf{k}(i)\mathbf{r}^{H}(i)\mathbf{R}^{-1}(i-1)\mathbf{r}(i).\label{eq:s2}
\end{split}
\end{equation}
Based on (\ref{eq:filter1}) and (\ref{eq:s2}) we have (\ref{eq:exp1}).

\subsection{Derivation of (\ref{eq:exp2})}
We note that
\begin{equation}
\begin{split}
\mathbf{R}(i)&=\lambda(i)\mathbf{R}(i-1)+\mathbf{r}(i)\mathbf{r}^{H}(i)
\\&=\mathbf{r}(i)\mathbf{r}^{H}(i)+\lambda(i)\mathbf{r}(i-1)\mathbf{r}^{H}(i-1)+\lambda(i)\lambda(i-1)\mathbf{r}(i-2)\mathbf{r}^{H}(i-2)\\&\quad+\cdots
+\lambda(i)\lambda(i-1)\cdots\lambda(2)\mathbf{r}(1)\mathbf{r}^{H}(1)+\lambda(i)\lambda(i-1)\cdots\lambda(2)\lambda(1)\mathbf{r}(0)\mathbf{r}^{H}(0).\label{eq:deassume1}
\end{split}
\end{equation}
%We assume that  forgetting factor $\gamma(m)$ and $\gamma(n)$ are uncorrelated, where $m, n=N_{i}\ldots i$, and . matrix $\mathbf{u}_{k}(n)\mathbf{u}^{H}_{k}(n)$ are uncorrelated,
%
For  large $i$, based on  (\ref{eq:ctvff1}), (\ref{eq:ctvff2}) and (\ref{eq:ctvff3}), we assume that there exists a number $N_{i}>0$, when $i\geq N_{i}$, for which we have that the forgetting factor $\lambda(i)$ varies slowly around its mean value and $E[\lambda(N_{i})]\approx E[\lambda(N_{i}+1)] \approx \cdots \approx E[\lambda(i)]\approx E[\lambda(\infty)]$.
%We make the following assumptions
%\begin{enumerate}[1)~]
%%$\textit{I})$
%\item
%Forgetting factor $\gamma(m)$ and $\gamma(n)$ are uncorrelated, where $m, n=N_{i}\ldots i$.
%% $\textit{II})$
%\item Forgetting factor $\gamma(i), \gamma(i-1),\ldots, \gamma(N_{i})$ are uncorrelated to the previous forgetting factors.
%%and compare it with our proposed low-complexity variable forgetting factor mechanism.
%\item
%Forgetting factor $\gamma(m)$ is uncorrelated to matrix $\mathbf{u}_{k}(n)\mathbf{u}^{H}_{k}(n)$,  where $m=N_{i}\ldots i$ and $n=N_{i}-1,\ldots, i-1$ .
%%Several chip-interleaving codebook design methods are proposed.
%\end{enumerate}

%
%$E[\gamma(1)\gamma(2)\cdots\gamma(i)]\approx \big( E[\gamma(i)]\big)^{i}$.
By taking the expectation of (\ref{eq:deassume1}) we obtain
\begin{equation}
\begin{split}
E[\mathbf{R}(i)]&\approx E[1+\lambda(i)+\lambda(i)\lambda(i-1)+\cdots
+\lambda(i)\lambda(i-1)\cdots\lambda(N_{i})]E[\mathbf{r}(i)\mathbf{r}^{H}(i)]\\&\quad+E[\lambda(i)\lambda(i-1)\cdots\lambda(N_{i})] E[\lambda(N_{i}-1)\mathbf{r}(N_{i}-2)\mathbf{r}^{H}(N_{i}-2)+\cdots\\&\quad+\lambda(N_{i}-1)\lambda(N_{i}-2)\cdots\lambda(2)\lambda(1)\mathbf{r}(0)\mathbf{r}^{H}(0)]
\\&\approx \big(1+E[\lambda(i)]+E^{2}[\lambda(i)]+\cdots+E^{i-N_{i}+1}[\lambda(i)])   E[\mathbf{r}(i)\mathbf{r}^{H}(i)]+E^{i-N_{i}+1}[\lambda(i)]\mbox{\boldmath$\Lambda$},
\end{split}
\end{equation}
where the matrix $\mbox{\boldmath$\Lambda$}=E[\lambda(N_{i}-1)\mathbf{r}(N_{i}-2)\mathbf{r}^{H}(N_{i}-2)+\cdots+\lambda(N_{i}-1)\lambda(N_{i}-2)\cdots\lambda(2)\lambda(1)\mathbf{r}(0)\mathbf{r}^{H}(0)]$ is determinate.
%, and we assumed $E[(\gamma(i))^{n}]\approx E^{n}[\gamma(i)]$.
Note that $0<E[\lambda(i)]<1$, when $i\rightarrow\infty$, we obtain
\begin{equation}
E[\mathbf{R}(i)]\approx \frac{1}{1-E[\lambda(\infty)]}E[\mathbf{r}(i)\mathbf{r}^{H}(i)].
\end{equation}
%
%By adjusting the parameter $\nu$ to have the convexity of the CCM design \cite{rcdl1, rcdl4} and using the expressions of the adaptive CCM-RLS receiver (\ref{eq:filter})-(\ref{eq:d}), when $i\rightarrow \infty$, due to the fact that the time average approximates the statistical ensemble average,
%we have $E[\mathbf{u}_{k}(i)\mathbf{u}^{H}_{k}(i)]=\mathbf{\bar{Q}}_{0}$ and
%\begin{equation}
%\begin{split}
%E^{-1}[\mathbf{R}(i)]&\approx (1-E[\lambda(\infty)])\mathbf{\bar{R}}^{-1}_{0}\\&\approx(1-E[\gamma(\infty)])E^{-1}[|z_{0}(i)|^{2}]\mathbf{R}^{-1}.
%\end{split}
%\end{equation}
%%Based on  \cite{tadali, eleftheriou, macchi},
%Note that, when $i\rightarrow \infty$, $\mathbf{Q}_{k}(i)$ and $\mathbf{Q}^{-1}_{k}(i)$  converge. Thus,
%we assume  $\lim_{i \rightarrow \infty}\mathbf{Q}^{-1}_{k}(i)\approx \lim_{i \rightarrow \infty} E[\mathbf{Q}^{-1}_{k}(i)]\approx \lim_{i \rightarrow \infty}E^{-1}[\mathbf{Q}_{k}(i)]$.
Note that, when $i\rightarrow \infty$, $\mathbf{R}(i)$ and $\mathbf{R}^{-1}(i)$  converge. Thus,
we assume  $\lim_{i \rightarrow \infty}\mathbf{R}^{-1}(i)\approx \lim_{i \rightarrow \infty} E[\mathbf{R}^{-1}(i)]\approx \lim_{i \rightarrow \infty}E^{-1}[\mathbf{R}(i)]$.
Finally, we obtain (\ref{eq:exp2}).

\end{appendix}

\end{document}